\documentclass[12pt]{article}
\usepackage{epsfig}
\usepackage{graphicx}
\usepackage{a4}
\usepackage{latexsym}
\usepackage{cite}

\usepackage{color}
\usepackage{colordvi}

\usepackage{pslatex}
\usepackage[latin1]{inputenc}
\usepackage[T1]{fontenc}

\newcommand{\as}{\alpha_{\rm s}}

\def\MSbar{\overline{\mathrm{MS}}}
\def\ep{\epsilon}
\def\z#1{{\zeta_{#1}}}
\def\ca{{C^{}_A}}
\def\cf{{C^{}_F}}
\def\tf{{T^{}_F}}
\def\nf{{n^{}_{\! f}}}
\def\nl{{n^{}_{\! l}}}
\def\nh{{n^{}_{\! h}}}

\def\lm{\mathrm{L}_m}
\def\ls{\mathrm{L}_s}
\def\lx{\mathrm{L}_x}
\def\ly{\mathrm{L}_y}
\def\s#1#2{\mathrm{S}_{#1,#2}(x)}
\def\li#1{\mathrm{Li}_{#1}(x)}
\newcommand{\ebrk}{\nonumber \\ &&}
\newcommand{\brk}{\right. \nonumber \\ && \left.}
\newcommand{\ibrk}{\right. \right. \nonumber \\ && \left. \left.}
\newcommand{\iibrk}{\right. \right. \right. \nonumber \\ && \left. \left. \left.}

\begin{document}
\setlength{\parskip}{0.2cm} \setlength{\baselineskip}{0.55cm}

\begin{titlepage}
\noindent
DESY 07-101 \hfill {\tt arXiv:0707.4139v1}\\
SFB/CPP-07-37 \\
July 2007 \\
\vspace{1.5cm}
\begin{center}
\LARGE {\bf
Heavy-quark production \\[0.5ex]
in gluon fusion at two loops in QCD
}\\
\vspace{2.2cm}
\large
M. Czakon$^{a,b}$, A. Mitov$^{c}$ and S. Moch$^{c}$ \\
\vspace{1.4cm}
\normalsize
{\it
$^{a}$Institut f\"ur Theoretische Physik und Astrophysik, Universit\"at W\"urzburg \\[0.5ex]
Am Hubland, D-97074 W\"urzburg, Germany \\[.5cm]
$^{b}$Department of Field Theory and Particle Physics,
Institute of Physics \\[0.5ex]
University of Silesia, Uniwersytecka 4, PL-40007 Katowice,
Poland \\[.5cm]
$^{c}$Deutsches Elektronensynchrotron DESY \\[0.5ex]
Platanenallee 6, D--15738 Zeuthen, Germany}
\vfill
\large {\bf Abstract}
\vspace{-0.2cm}
\end{center}
We present the two-loop virtual QCD corrections to the production of
heavy quarks in gluon fusion.
The results are exact in the limit when all kinematical invariants are large
compared to the mass of the heavy quark up to terms suppressed by powers
of the heavy-quark mass.
Our derivation uses a simple relation between massless and massive QCD scattering amplitudes
as well as a direct calculation of the massive amplitude at two loops.
The results presented here together with those obtained previously for quark-quark scattering
form important parts of the next-to-next-to-leading order QCD corrections to
heavy-quark production in hadron-hadron collisions.
\\
\vspace{2.0cm}
\end{titlepage}

\newpage

%
%
\section{Introduction}
\label{sec:intro}

The production of heavy quarks at hadron colliders is an important process for a range of reasons.
Experimentally, events with a top-quark, being the heaviest quark known thus far, lead to very
characteristic signatures in a hadronic scattering process.
This allows for event reconstruction in a variety of channels, which e.g. at LHC is supplemented
by an anticipated very high statistics for the production cross section~\cite{atlas:1999tdr,cms:2006tdr}.
In this way precision measurements of the top-mass, of (differential) $t {\bar t}$-distributions
and also tests of the production and the subsequent decay mechanism
(including anomalous couplings and top-spin correlations) can be conducted at Tevatron or LHC.
The precise knowledge of the top-quark parameters has direct impact on
precision tests of the Standard Model, in particular on the Higgs sector due to the large top-Yukawa coupling.
Moreover, events with top-quarks will also make up for a large part of the background in
searches for the Higgs boson or new physics.

In the case of bottom-quark production important measurements are inclusive $B$-meson and 
$b$-jet production at moderate to large transverse momentum.
At colliders $b$-flavored jets are produced for instance in decays of top-quarks, the Higgs boson
and numerous particles proposed in extensions of the Standard Model.
Differential $b$-jet distributions are also important for measurements of parton distributions.
Predictions for heavy-quark hadro-production have theoretical uncertainties,
which can even be larger than e.g. the corresponding predictions for the production of light jets.
This is due to radiative corrections in Quantum Chromodynamics (QCD),
where typically large logarithms at higher orders appear.
At present, the next-to-leading order (NLO) QCD corrections for
heavy-quark hadro-production are
known~\cite{Nason:1988xz,Nason:1989zy,Beenakker:1989bq,Beenakker:1991ma,Mangano:1992jk,Korner:2002hy,Bernreuther:2004jv}
and, once they become available, the complete next-to-next-to-leading order (NNLO) QCD corrections will
reduce the uncertainty of theory predictions by improving the stability with respect to scale variations.
For inclusive $B$-meson production~\cite{Cacciari:2002pa,Cacciari:2003uh}, for instance,
the formalism of perturbative fragmentation functions of a heavy quark~\cite{Mele:1990cw}
has already been extended through NNLO by providing the initial conditions~\cite{Melnikov:2004bm,Mitov:2004du}
together with the time-like (non-singlet) evolution~\cite{Mitov:2006ic}.
Alternative ways devoted to control the theoretical uncertainty in the case of $b$-quarks,
for example through the definition of dedicated jet algorithms have been
proposed recently~\cite{Banfi:2007gu}.

In the hadronic production of heavy quarks large perturbative QCD corrections
arise in different kinematical regions.
If the heavy quarks are produced with (partonic) center-of-mass energies $s$ close to threshold,
then Sudakov-type logarithms, powers of $\log(\beta)$,
appear to all orders in perturbation theory,
where $\beta=\sqrt{1-4m^2/s}$ is the velocity of the heavy quark.
The necessary resummation has been carried out
to next-to-leading logarithmic (NLL) accuracy~\cite{Kidonakis:1997gm,Bonciani:1998vc,Kidonakis:2001nj}
and has successfully improved the phenomenology of top-quark production at Tevatron.
In the high-energy regime on the other hand, when the heavy quarks are fast,
logarithms in the quark mass $m$ appear. These $\log(m)$-terms are of
collinear origin and dominate cross sections when $m$
becomes negligible in comparison to other kinematical invariants, such as
e.g. for the inclusive $b$-jet spectrum at large transverse momentum.
All-order predictions can be achieved by means of an explicit resummation.
The necessary technology for resumming both the incoming and outgoing collinear
logarithms to NLL accuracy is available,
see e.g. Refs.~\cite{Ball:2001pq,Marchesini:2004ne} and references therein.

In this article, we want to further improve the precision of fixed-order perturbative QCD corrections.
To that end, we present results for the virtual QCD corrections
at two loops for the pair-production of heavy quarks in gluon fusion.
Together with the corresponding results for quark-quark scattering obtained
previously by us~\cite{Czakon:2007ej} these are essential parts of the complete
NNLO QCD corrections.
To be precise, we calculate the interference of the two-loop with the Born amplitude.
We work in the limit of fixed scattering angle and high energy,
where all kinematic invariants are large compared to the heavy-quark mass $m$.
Thus, our result contains all logarithms $\log(m)$ as well
as all constant contributions (i.e. the mass-independent terms) and
we consistently neglect power corrections in $m$.

In our calculation we employ two different methods.
On the one hand, we apply a generalization of the infrared factorization formula for
massless QCD amplitudes~\cite{Catani:1998bh,Sterman:2002qn} to the
case of massive partons~\cite{Mitov:2006xs}.
In a nut-shell this results in an extremely simple universal multiplicative relation
between a massive QCD amplitude in the small-mass limit and its
massless version~\cite{Mitov:2006xs}.
In this way, we can largely use for our derivation the results
of the NNLO QCD corrections to massless quark-gluon scattering (i.e. $g g \to q {\bar q}$).
These have been computed for the squared matrix element,
i.e. the interference of the two-loop virtual corrections with the Born
amplitude in Ref.~\cite{Anastasiou:2001sv}, while
results for the individual (independent) helicity configurations
of the two-loop amplitude $|{\cal M}^{(2)} \rangle$ itself
have been given in Refs.~\cite{Bern:2003ck,Glover:2003cm}.
On the other hand, we perform a direct calculation of the relevant
Feynman diagrams in the massive case followed by a subsequent
expansion in the small-mass limit by means of Mellin-Barnes representations.

The outline of the article is as follows.
In Section~\ref{sec:setstage} we give some basic definitions and
present a short summary of our methods in Section~\ref{sec:method}.
There we briefly explain the essence of the QCD factorization approach to
calculate massive amplitudes and give the relevant formulae.
We also highlight the key steps of the direct evaluation of Feynman diagrams
in the small-mass limit and, in particular, comment on the non-planar topologies.
Section~\ref{sec:results} contains our results and we conclude in Section~\ref{sec:conclusions}.
Appendix~\ref{sec:appA} gives the explicit result for a massive non-planar scalar integral
in the small mass limit.

%
%
\section{Setting the stage}
\label{sec:setstage}

The pair-production of heavy quarks in the gluon fusion process
corresponds to the scattering
\begin{equation}
\label{eq:ggQQ}
g(p_1) + g(p_2) \:\:\rightarrow\:\: Q(p_3,m) + {\bar Q}(p_4,m) \, ,
\end{equation}
where $p_i$ denote the gluon and quark momenta and $m$ the mass of the heavy quark.
Energy-momentum conservation implies
\begin{equation}
\label{eq:engmom}
p_1^\mu+p_2^\mu = p_3^\mu+p_4^\mu \, .
\end{equation}
Following the notation of Ref.~\cite{Anastasiou:2001sv} we consider
the scattering amplitude ${\cal M}$ for the process~(\ref{eq:ggQQ})
at fixed values of the external parton momenta $p_i$, thus $p_1^2 =
p_2^2 = 0$ and $p_3^2 = p_4^2 = m^2$.
The amplitude ${\cal M}$ may be written as a series expansion in the strong coupling $\as$,
\begin{eqnarray}
  \label{eq:Mexp}
  | {\cal M} \rangle
  & = &
  4 \pi \as \biggl[
  | {\cal M}^{(0)} \rangle
  + \biggl( {\as \over 2 \pi} \biggr) | {\cal M}^{(1)} \rangle
  + \biggl( {\as \over 2 \pi} \biggr)^2 | {\cal M}^{(2)} \rangle
  + {\cal O}(\as^3)
  \biggr]
\, ,
\end{eqnarray}
and we define the expansion coefficients in powers of $\as(\mu^2)
/ (2\pi)$ with $\mu$ being the renormalization scale. We work in
conventional dimensional regularization, $d=4-2 \ep$, in the
$\MSbar$-scheme for the coupling constant renormalization. The heavy
mass $m$ on the other hand is always taken to be the pole mass.

We explicitly relate the bare (unrenormalized) coupling $\as^{\rm{b}}$
to the renormalized coupling $\as$ by
\begin{eqnarray}
\label{eq:alpha-s-renorm}
\as^{\rm{b}} S_\epsilon \: = \: \as
\biggl[
   1
   - {\beta_0 \over \epsilon} \biggl( {\as \over 2 \pi} \biggr)
   + \left(
     {\beta_0^2 \over \epsilon^2}
     - {1 \over 2} {\beta_1 \over \epsilon}
     \right) \biggl( {\as \over 2 \pi} \biggr)^2
  + {\cal O}(\as^3)
  \biggr]
\, ,
\end{eqnarray}
where we put the factor $S_\epsilon=(4 \pi)^\ep \exp(-\ep \gamma_{\rm E}) = 1$
for simplicity and $\beta$ is the QCD $\beta$-function
\cite{van Ritbergen:1997va,Czakon:2004bu}
\begin{eqnarray}
\label{eq:betafct}
\beta_0 = {11 \over 6}\*\ca - {2 \over 3}\*\tf\*\nf \, ,
\qquad
\beta_1 = {17 \over 6}\*\ca^2 - {5 \over 3}\*\ca\*\tf\*\nf - \cf\*\tf\*\nf \, .
\end{eqnarray}
The color factors are in a non-Abelian ${\rm{SU}}(N)$-gauge theory
$\ca = N$, $\cf = (N^2-1)/2\*N$ and $\tf = 1/2$.
Throughout this letter, $N$ denotes the number of colors and
$\nf$ the total number of flavors, which is the sum of
$\nl$ light and $\nh$ heavy quarks.

In the following, we will confine ourselves to the discussion of the squared amplitude for the process~(\ref{eq:ggQQ}),
although it should be clear that our approach and the results of the present article can be easily extended to the
(color ordered) partial amplitudes for the individual helicity combinations of the massive two-loop amplitude
$|{\cal M}^{(2)} \rangle$ itself. There one would rely in particular on Refs.~\cite{Bern:2003ck,Glover:2003cm}.

For convenience, we define the function ${\cal A}(\epsilon, m, s, t, \mu)$
for the squared amplitudes summed over spins and colors as
\begin{eqnarray}
\label{eq:Msqrd}
\overline{\sum |{\cal M}({g + g \to  Q + \bar{Q}} )|^2}
&=&
{\cal A}(\epsilon, m, s, t, \mu)
\, .
\end{eqnarray}
${\cal A}$ is a function of the Mandelstam variables $s$, $t$ and $u$ given by
\begin{equation}
\label{eq:Mandelstam}
s = (p_1+p_2)^2\, , \qquad
t  = (p_1-p_3)^2 - m^2\, , \qquad
u  = (p_1-p_4)^2 - m^2\, ,
\end{equation}
and has a perturbative expansion similar to Eq.~(\ref{eq:Mexp}),
\begin{equation}
\label{eq:Aexp}
{\cal A}(\epsilon, m, s, t, \mu) = 16 \pi^2 \as^2
\left[
  {\cal A}^4
  + \biggl( {\as \over 2 \pi} \biggr) {\cal A}^6
  + \biggl( {\as \over 2 \pi} \biggr)^2 {\cal A}^8
  + {\cal O}(\as^{3})
\right]
\, .
\end{equation}
In terms of the amplitudes the expansion coefficients in Eq.~(\ref{eq:Aexp})
may be expressed as
\begin{eqnarray}
\label{eq:A4def}
{\cal A}^4 &=&
\langle {\cal M}^{(0)} | {\cal M}^{(0)} \rangle \equiv
2{(N^2-1) \over N}(1-\epsilon)\left({N^2-1 \over ut} - 2 {N^2 \over s^2} \right)
\left(t^2 + u^2 - \epsilon s^2\right) + {\cal O}(m)\, , \\
\label{eq:A6def}
{\cal A}^6 &=& \left(
\langle {\cal M}^{(0)} | {\cal M}^{(1)} \rangle + \langle {\cal M}^{(1)} | {\cal M}^{(0)} \rangle
\right)\, , \\
\label{eq:A8def}
{\cal A}^8 &=& \left(
\langle {\cal M}^{(1)} | {\cal M}^{(1)} \rangle
+ \langle {\cal M}^{(0)} | {\cal M}^{(2)} \rangle + \langle {\cal M}^{(2)} | {\cal M}^{(0)} \rangle
\right)\, ,
\end{eqnarray}
where we have discarded powers in the heavy-quark mass $m$ in ${\cal A}^4$.

The expressions for ${\cal A}^6$ have been presented e.g. in Refs.~\cite{Korner:2002hy,Bernreuther:2004jv} and
the loop-by-loop contribution $\langle {\cal M}^{(1)} | {\cal M}^{(1)} \rangle$ at NNLO in ${\cal A}^8$
has been published in Ref.~\cite{Korner:2005rg}.
Both results for ${\cal A}^6$ and $\langle {\cal M}^{(1)} | {\cal M}^{(1)}
\rangle$ have been obtained in dimensional regularization and with the complete dependence on the heavy-quark mass.
Here we provide for the first time the real part of
$\langle {\cal M}^{(0)} | {\cal M}^{(2)} \rangle$ up to powers ${\cal O}(m)$ in the heavy-quark mass $m$.

%
%
\section{Method}
\label{sec:method}

%
%
\subsection{The massive amplitude from QCD factorization}
Let us briefly recall the key findings of Ref.~\cite{Mitov:2006xs}
on how to calculate loop amplitudes with massive partons from purely
massless amplitudes. Heuristically, the QCD factorization approach
rests on the fact that a massive amplitude ${\cal M}^{(m)}$
for any given physical process shares essential properties in the
small-mass limit with the corresponding massless amplitude ${\cal
M}^{(m=0)}$. The latter one, ${\cal M}^{(m=0)}$,
generally displays two types of singularities, soft and collinear,
related to the emission of gluons with vanishing energy and to
collinear parton radiation off massless hard partons, respectively.
These appear explicitly as poles in $\ep$ in dimensional
regularization after the usual ultraviolet renormalization is
performed. In the former case, the soft singularities remain in
${\cal M}^{(m)}$ as single poles in $\ep$ while some of the
collinear singularities are now screened by the mass $m$ of the
heavy fields, which gives rise to a logarithmic dependence on $m$,
see e.g. Ref.~\cite{Catani:2000ef}.

This structure of singularities for massless amplitudes
has been clarified to all orders in perturbation theory~\cite{Sterman:2002qn}
as all $1/\epsilon$ terms can be exponentiated, see also Ref.~\cite{MertAybat:2006mz}.
Similarly, all poles in $\epsilon$ and $\log(m)$ terms for amplitudes with
massive partons also obey an all-order exponentiation
with mostly the same anomalous dimensions as in the massless case~\cite{Mitov:2006xs}.
Thus, in the small-mass limit the differences between a massless and
a massive amplitude can be thought of as due to the difference in
the infrared regularization schemes. QCD factorization provides a
remarkably simple direct relation between ${\cal M}^{(m)}$
and ${\cal M}^{(m=0)}$
\begin{eqnarray}
\label{eq:Mm-M0}
{\cal M}^{(m)} &=&
\prod_{i\in\ \{{\rm all}\ {\rm legs}\}}\,
  \left(
    Z^{(m\vert0)}_{[i]}
  \right)^{1 \over 2}\,
  \times\
{\cal M}^{(m=0)}\, .
\end{eqnarray}
The function $Z^{(m\vert 0)}$ is process independent and depends
only on the type of external parton, i.e. quarks and gluons in the
case at hand. For external massive quarks $Q$ it is defined as the
ratio of the on-shell heavy-quark form factor and the massless
on-shell one, both being
known~\cite{Bernreuther:2004ih,Moch:2005id,Gehrmann:2005pd} to
sufficient orders in $\as$ and powers of $\ep$. An explicit
expression for
\begin{equation}
\label{eq:ZQ}
Z^{(m\vert0)}_{[Q]} \, = \, 1 +
\sum\limits_{j=1}^{\infty} \left( {\as \over 2 \pi} \right)^j\,
Z^{(j)}_{[Q]} \, ,
\end{equation}
up to two loops is given in Ref.~\cite{Mitov:2006xs} (note the
different normalization ${\as /  (4 \pi)}$ used there).
The leading $\nf$ terms $\sim(\nf\as)^n$ for our process Eq.~(\ref{eq:ggQQ}),
$gg\to Q{\bar Q}$,
can also be predicted based on the above arguments.
Keeping only terms quadratic in $\nh$ and/or $\nf = \nh+\nl$ one
has up to two loops:
\begin{equation}
\label{eq:Zg}
Z^{(m\vert0)}_{[g]} \, = \, 1 + {\as \over 2 \pi} \,
Z^{(1)}_{[g]} + \left( {\as \over 2 \pi} \right)^2\, Z^{(2)}_{[g]}
\, +{\cal O}(\as^3)\, ,
\end{equation}
where
\begin{equation}
\label{eq:Zgtwo}
Z^{(2)}_{[g]}\, =\, \left( Z^{(1)}_{[g]} \right)^2
+\, {2\over 3\ep}\,\nf\tf\, Z^{(1)}_{[g]} + {\cal O}(\nh^1 \times \nl^0)\, ,
\end{equation}
with $Z^{(1)}_{[g]} \sim \nh$ given in Ref.~\cite{Mitov:2006xs}
(again note the different normalization ${\as /  (4 \pi)}$ used there).
$Z^{(1)}_{[g]}$ is also equal to the ${\cal O}(\as)$ term in the gluon wave
function renormalization constant $Z_3$ in Eq.~(\ref{eq:Z3}). 
The relation of $Z^{(m\vert0)}_{[g]}$ to $Z_3$ is discussed after
Eq.~(\ref{eq:Z3ren}) below. To derive Eq.~(\ref{eq:Zgtwo}) we apply the
definition for $Z^{(m\vert 0)}$ given in Ref.~\cite{Mitov:2006xs},
i.e. evaluate the ratio of the gluon form factor with heavy-loop
insertions and the pure massless gluon form
factor~\cite{Gehrmann:2005pd,Moch:2005tm}. The additional
renormalization constant that enters the effective $Hgg$ vertex (see
e.g. Ref.~\cite{Moch:2005tm} for details) cancels in the ratio and
does not contribute to $Z^{(m\vert0)}_{[g]}$. Exploiting the
predictive power of the relation Eq.~(\ref{eq:Mm-M0}) and applying
it to the process Eq.~(\ref{eq:ggQQ}) we get
\begin{eqnarray}
\label{eq:A8mtoA80}
2 {\rm Re}\, \langle {\cal M}^{(0)} | {\cal M}^{(2)} \rangle^{(m)}
&=& 2 {\rm Re}\, \langle {\cal M}^{(0)} |
{\cal M}^{(2)} \rangle^{(m=0)}\ +\ \left(Z^{(1)}_{[Q]} +
Z^{(1)}_{[g]}\right) {\cal A}^{6,(m=0)}\nonumber \\
&+&  2 \left(Z^{(2)}_{[Q]} + Z^{(2)}_{[g]} + Z^{(1)}_{[Q]}
Z^{(1)}_{[g]}\right) {\cal A}^{4,(m=0)}\
+\ {\cal O}(\nh^1 \times \nl^0) +\ {\cal O}(m)
\, ,
\end{eqnarray}
which assumes the hierarchy of scales $m^2 \ll s,t,u$ , i.e. we
neglect terms ${\cal O}(m)$. Eq.~(\ref{eq:A8mtoA80}) predicts the
complete real part of the squared amplitude
$\langle {\cal M}^{(0)} | {\cal M}^{(2)} \rangle^{(m)}$
except (as indicated) for those terms, which are linear in $\nh$ (the
number of heavy quarks) and, at the same time not proportional to $n_l$ (the
number of light quarks).
These two-loop contributions have been excluded explicitly also from the definition~\cite{Mitov:2006xs}
of $Z^{(m\vert 0)}$, as one needs additional process dependent terms for their description.

The two-loop massless amplitudes
${\rm Re}\, \langle {\cal M}^{(0)} | {\cal M}^{(2)} \rangle^{(m=0)}$ are computed
in Ref.~\cite{Anastasiou:2001sv}.
We have checked that the finite remainders of the squared two-loop amplitudes obtained after
the infrared subtraction procedure discussed in that reference agree with the
corresponding terms constructed from the two-loop helicity
amplitudes calculated in Ref.~\cite{Bern:2003ck}. We have also found
similar agreement between the finite remainders of the
$q{\bar q} \to q'{\bar q}'$ amplitudes we used in Ref.~\cite{Czakon:2007ej}
that we extracted from Refs.~\cite{Anastasiou:2000kg,DeFreitas:2004tk}.

%
%
\subsection{Direct computation of the massive amplitude}
The direct computation of the massive amplitude proceeds
according to the same scheme as in our previous publication~\cite{Czakon:2007ej},
which itself was an evolution of the methodology developed
in Refs.~\cite{Czakon:2004wm,Czakon:2006pa,Actis:2007gi}.
In short, the complete amplitude is reduced to an expression containing only a small number
of integrals with the help of the Laporta algorithm~\cite{Laporta:2001dd}. In
a next step, Mellin-Barnes (MB) representations~\cite{Smirnov:1999gc,Tausk:1999vh}
of all these integrals are constructed, and
analytically continued in the dimension of space-time with the help of the
{\tt MB} package~\cite{Czakon:2005rk} revealing the full singularity
structure. A subsequent asymptotic expansion in the mass parameter is done by
closing contours and resumming the integrals, either with the help of {\tt XSummer}~\cite{Moch:2005uc},
or the {\tt PSLQ} algorithm~\cite{pslq:1992}.

\begin{figure}[ht]
  \parbox[t]{7.5cm}{\center
    \parbox[c][5cm][c]{7cm}{\center
      \epsfig{file=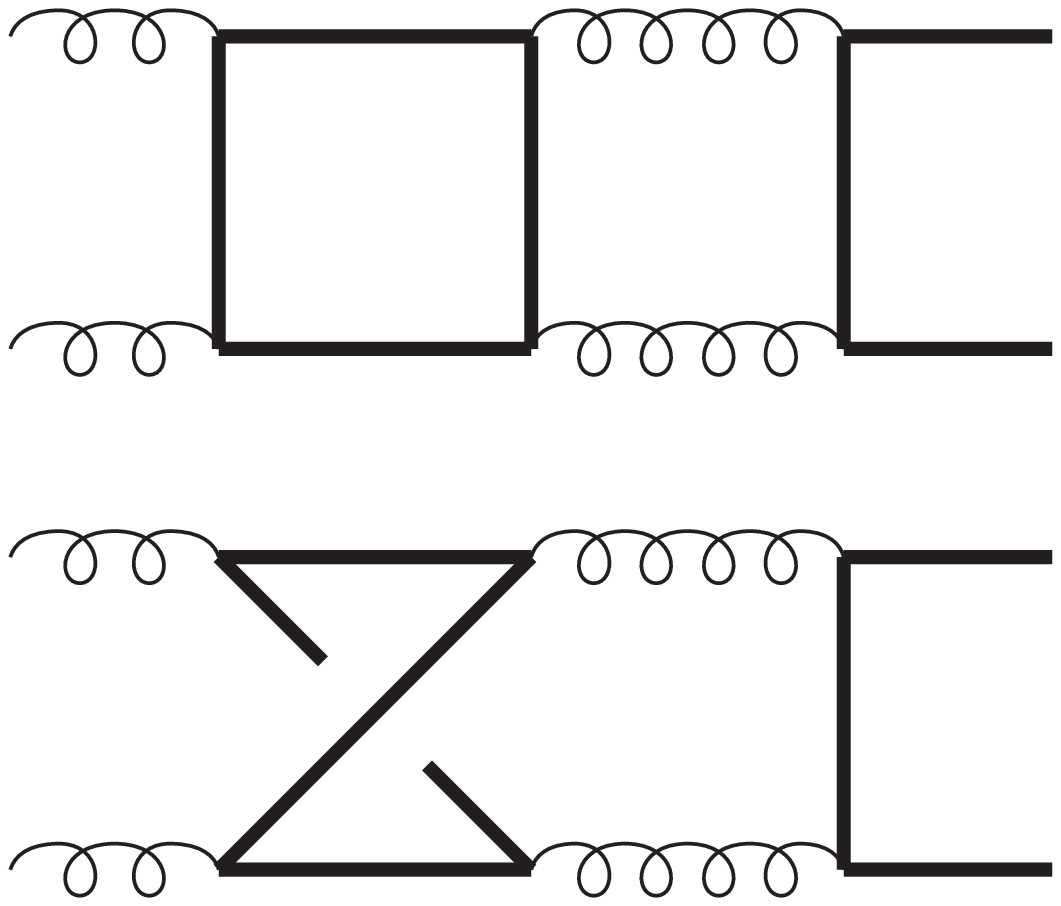,width=5cm}}
    \caption{\label{fig:diags}
      Most complicated diagrams of the single heavy quark loop
      contribution. The thick lines are massive.
    }
    \parbox[c][4cm][c]{7cm}{\center
      \epsfig{file=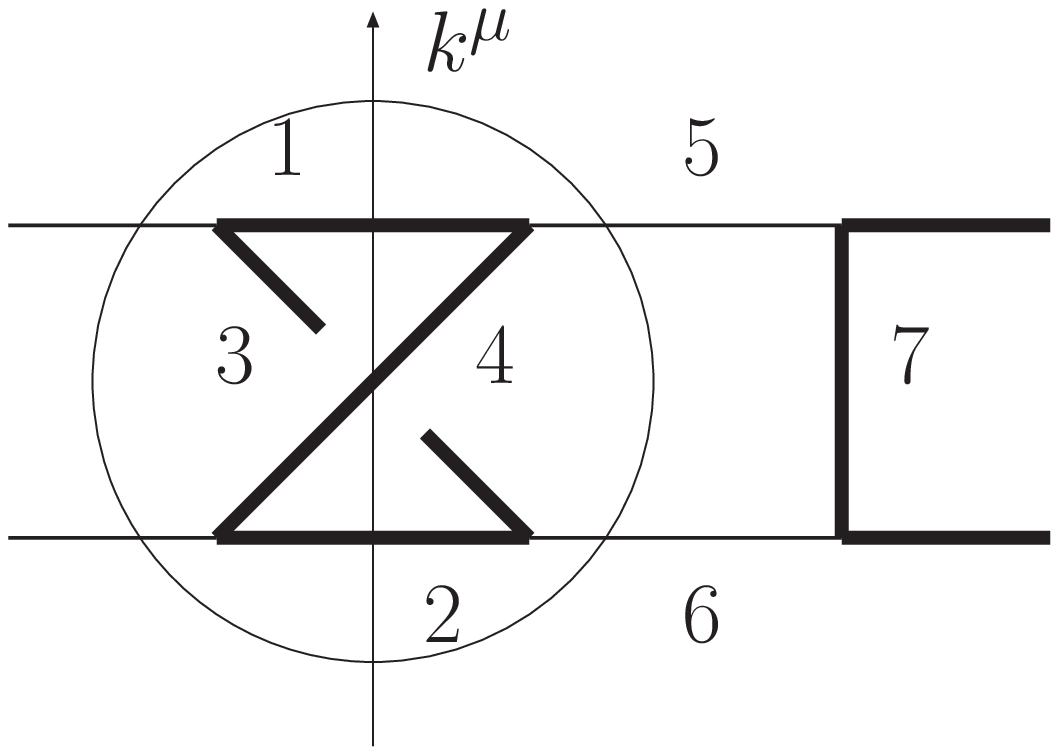,width=5cm}}
    \caption{\label{fig:npnum}
      Definition of the momentum in the numerator of the non-planar integral
      considered in the text, together with the labeling of the denominators.
    }
  }
  \hspace{1cm}
  \parbox[t]{7.5cm}{\center \parbox[c][7.3cm][c]{8cm}{\center
    \epsfig{file=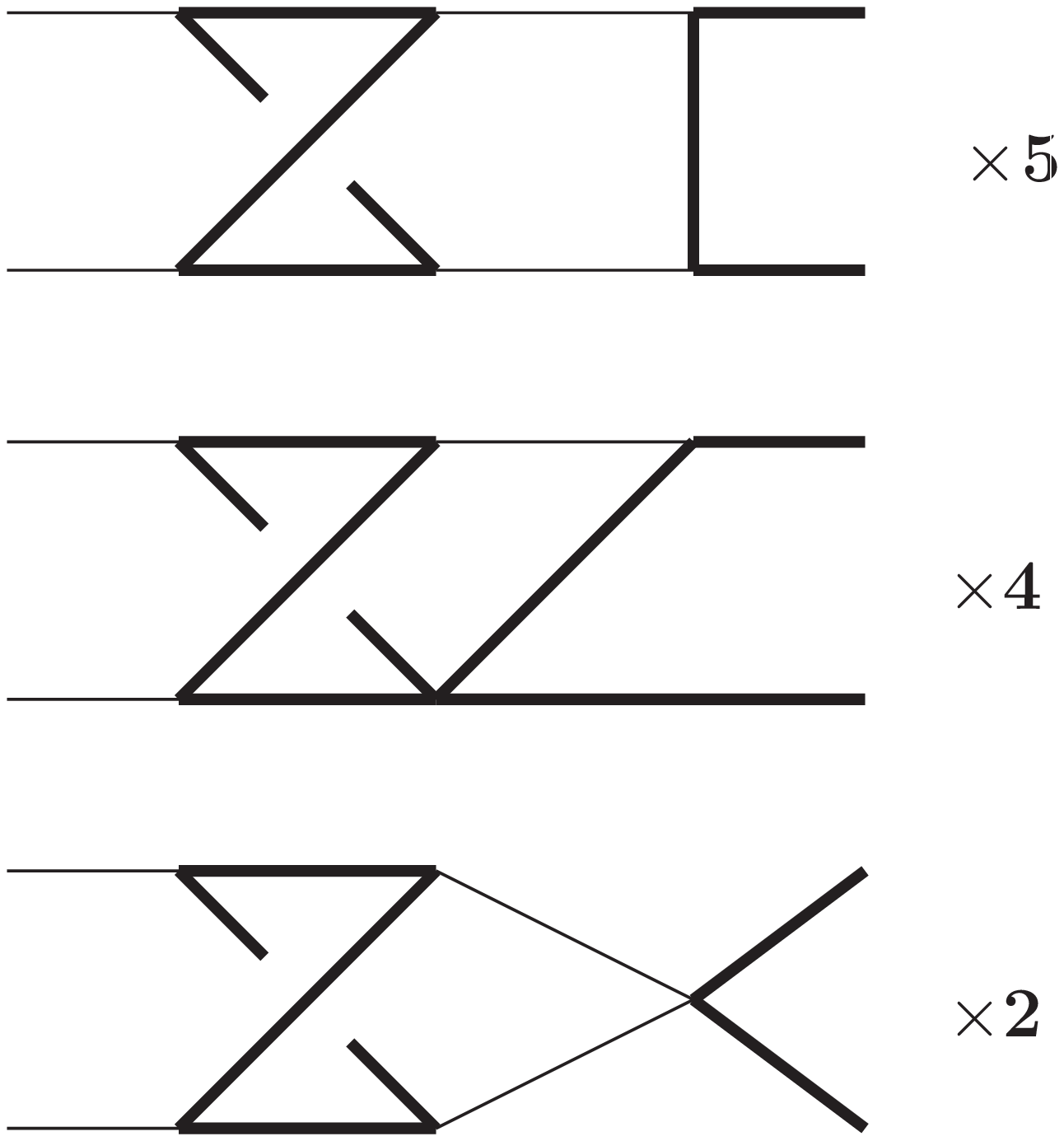,width=6cm}}
    \caption{\label{fig:ints}
    Non-planar master integrals corresponding to the second diagram of
    Fig.~\ref{fig:diags}.  The numbers denote the multiplicities of the
    integrals of the given topology. All of the required MB representations
    can be derived from the MB representation of the integral Fig.~\ref{fig:npnum}.
    }}
\end{figure}

We shall now concentrate on the differences with respect to our previous
calculation. First of all, the number of master integrals is substantially
larger, reaching 422, which adds a lot to the complexity of the
computation. This is partly due to the fact, that the symmetry with
respect to the exchange of the gluons generates the same topologies in the $t$-
and $u$-channels, but more importantly because of completely new
topologies, which come together with the more extended set of gluon
interactions. Whereas in Ref.~\cite{Czakon:2007ej}, it was possible to
avoid the computation of the high-energy asymptotics of non-planar graphs, and
still have a test of the factorization approach, we were not able to avoid
them here. In fact, this additional complication is due to the single heavy-quark
loop diagrams of Fig.~\ref{fig:diags}, which are explicitely removed
from the factorization approach. The complete set of non-planar master
integrals belonging to this class is depicted in Fig.~\ref{fig:ints}.

The first problem that one has to face when dealing with non-planar integrals
is the construction of MB representations. In the planar case, the iterative
loop-by-loop integration has proved to be the most fruitful. On the other
hand, the first non-planar double-box diagram ever
computed~\cite{Tausk:1999vh}, with massless propagators and on-shell external legs, had
its four-dimensional MB representation, $I_{NP}^{m=0,\;dim=4}$, derived directly
from the two-loop Feynman parameter
representation. It seems, however, that when masses are involved, the
loop-by-loop representations are more compact, as seen for example in
Ref.~\cite{Smirnov:2004ip,Heinrich:2004iq}. However, any asymptotic expansion
contains a so-called hard part, which is obtained by setting all the small
parameters to zero, and which in our case would correspond to the massless graph.
Following this line of thought, one can derive a representation for the massless
on-shell graph by taking suitable residues in the result presented in
Ref.~\cite{Smirnov:2004ip}. Unfortunately, one arrives at a six-fold
representation, $I_{NP}^{m=0,\;dim=6}$, in clear disadvantage with respect to
$I_{NP}^{m=0,\;dim=4}$. Interestingly, there is an even more severe problem
inherent in the loop-by-loop approach. The leading pole derived in
Ref.~\cite{Tausk:1999vh} reads (up to normalization factors irrelevant for
this discussion)
\begin{equation}
I_{NP}^{m=0,\;dim=4} =
  \frac{2}{\ep^4 s t u} + {\cal O}\left(\frac{1}{\ep^3}\right),
\end{equation}
whereas the six-fold representation gives (with the same normalization)
\begin{equation}
I_{NP}^{m=0,\;dim=6} =
  \frac{5}{2 \ep^4 s t u} + {\cal O}\left(\frac{1}{\ep^3}\right).
\end{equation}
This clear discrepancy is only explained when we look at the subleading pole
from the six-fold representation, which contains a logarithmic singularity
\begin{equation}
I_{NP}^{m=0,\;dim=6} \mid_{u \rightarrow -s-t} \; \approx \;
-\frac{1}{\ep^3 s t u} \log (-s-t-u) + {\cal O}\left(\frac{1}{\ep^2}\right).
\end{equation}
Obviously, the extension of the integral into the Euclidean domain, performed
with the help of the $u$ parameter, regularized part of the infrared
singularity. The only way to obtain the correct result would be to first take
the limit $u \rightarrow -s-t$, and only then $\ep \rightarrow 0$. This,
however, is a highly non-trivial task, as is well known from studies aiming at
the derivation of exact expressions of Feynman integrals in $d$-dimensions.

In view of all the above arguments, we derived our MB representations directly
from the two-loop Feynman parametric representation. In particular, for the
scalar integral of Fig.~\ref{fig:npnum}, we have the following six-fold
representation
\begin{eqnarray}
  \label{rep1}
I_{NP} &=&
  -(-s)^{-2 {\ep}-3}
  \int_{-i \infty}^{i \infty} \; \prod_{i=1}^{6} \; dz_i \;
  \left(-\frac{s}{{m^2}}\right)^{{z_1}}
  \left(-\frac{t}{{m^2}}\right)^{{z_2}}
  \left(-\frac{u}{{m^2}}\right)^{{z_3}}
  \ebrk \times
  \Gamma(-{z_2}) \Gamma(-{z_3}) \Gamma (-{z_4}) \Gamma(-{z_5}) \Gamma(-{z_6})
  \Gamma(-2 {z_1}-{z_2}-{z_3}-2 {z_4}+1)
  \Gamma(-{\ep}-{z_4})^2
  \ebrk \times
  \Gamma({z_1}+{z_2}+{z_3}+{z_4})
  \Gamma(-{\ep}+{z_1}+{z_2}+{z_3}+{z_4}-{z_5}-1)
  \Gamma(-2 {\ep}+{z_1}-{z_6}-2)
  \ebrk \times
  \Gamma(-2 {\ep}+{z_1}+{z_2}-{z_5}-{z_6}-2)
  \Gamma(-2 {\ep}+{z_1}+{z_3}-{z_5}-{z_6}-2)
  \ebrk \times
  \Gamma({z_2}+{z_6}+1)
  \Gamma({z_3}+{z_6}+1)
  \Gamma({z_5}+{z_6}+1)
  \Gamma(2 {\ep}-{z_1}+{z_5}+{z_6}+3)
  \ebrk \times
  \biggl(\Gamma(-2 {\ep}-2 {z_4})
  \Gamma(-3 {\ep}-{z_4}-1)
  \Gamma(-2 {\ep}+{z_1}+{z_2}+{z_3}-{z_5}-1)^2\biggr)^{-1},
\end{eqnarray}
where the loop integration is done with the measure $e^{\ep \gamma_E} \int d^d k/(i\pi^{d/2})$ per loop.
We defer the presentation of the full result of the expansion of this integral to the
Appendix~\ref{sec:appA}. Here we only note, that the leading term of the
expansion has a square root singularity, which is a feature of non-planar
graphs, that does not occur in any of the planar integrals considered in this
calculation. Clearly, the disappearance of this square root singularity is a
simple test of the correctness of the calculation.

The second problem that requires care is connected with the choice of the master
integrals. In Fig.~\ref{fig:ints}, we have not only shown the topologies, but
also the multiplicities of the masters. The basic seven-liner needs as much
as five different integrals.
It is clear that we want to avoid coefficients containing poles in $\ep$ or $m^2$,
since the leading behavior of the integrals is difficult enough to determine,
and such poles would be synonymous of higher orders in the respective expansions.
After inspection it turns out that one can take two integrals with second powers of
the denominators into the set, but tensors rank one and two are
unavoidable. Although it is possible to generate representations for arbitrary
tensors, one ends up with a huge number of four-fold integrals after
expansion. Instead, one can introduce a new ficticious propagator that will
have a negative power. For this we choose the square of the momentum that runs
through the crossed box subloop, as shown in Fig.~\ref{fig:npnum}.
\begin{figure}[ht]
  \begin{center}
    \epsfig{file=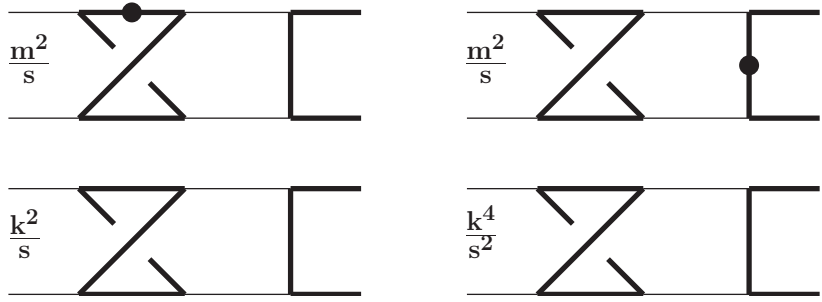,width=11cm}
    \caption{\label{fig:masters}
      A suitable choice of integrals that forms together with the underlying scalar
      integral the set of five master integrals of the seven line
      non-planar topology. The momentum $k^\mu$ has been defined in
      Fig.~\ref{fig:npnum}, whereas a dot on a line denotes a squared propagator.
    }
  \end{center}
\end{figure}

The final set of seven-line non-planar master integrals is shown in
Fig.~\ref{fig:masters}. Note that the dotted masters, i.e. those having higher
powers of chosen propagators are particularly easy to calculate, because one
only needs the singularities in $1/m^2$. All of the representations can be
derived from the following one
\begin{eqnarray}
  \label{rep2}
\overline{I}_{NP}  &=& (-1)^a (-s)^{-a-2 {\ep}+4}
  \int_{-i \infty}^{i \infty} \; \prod_{i=1}^{8} \; \frac{dz_i}{\Gamma(a_i)} \;
  \left(-\frac{s}{{m^2}}\right)^{{z_1}}
  \left(-\frac{t}{{m^2}}\right)^{{z_2}}
  \left(-\frac{u}{{m^2}}\right)^{{z_3}}
  \ebrk \times
  \Gamma(-{z_3})
  \Gamma(-{z_4})
  \Gamma(-{z_5})
  \Gamma(-{z_6})
  \Gamma(-{z_7})
  \Gamma(-{z_8})
  \Gamma(-{z_2}+{z_4}+{z_5})
  \Gamma({z_1}+{z_2}+{z_3}+{z_6})
  \ebrk \times
  \Gamma(a+2 {\ep}-{z_1}+{z_7}+{z_8}-4)
  \Gamma({z_2}+{z_3}-{z_4}-{z_5}+{z_8}+a_{1})
  \Gamma({z_5}+{z_8}+a_{4})
  \Gamma({z_7}+{z_8}+a_{6})
  \ebrk \times
  \Gamma(-2 {z_1}-2 {z_2}-2 {z_3}+{z_4}+{z_5}-2 {z_6}+a_{7})
  \Gamma({z_2}-{z_5}+a_{8})
  \Gamma(-{\ep}-{z_6}-a_{13}+2)
  \ebrk \times
  \Gamma(-{\ep}-{z_6}-a_{24}+2)
  \Gamma(-2 {z_1}-{z_2}-{z_3}+{z_4}-2 {z_6}+a_{78})
  \ebrk \times
  \Gamma(-{\ep}+{z_1}+{z_2}+{z_3}-{z_4}+{z_6}-{z_7}-a_{5678}+2)
  \Gamma(-2 {\ep}+{z_1}-{z_8}-a_{1234678}+4)
  \ebrk \times
  \Gamma(-2 {\ep}+{z_1}+{z_5}-{z_7}-{z_8}-a_{1245678}+4)
  \ebrk \times
  \Gamma(-2 {\ep}+{z_1}+{z_2}+{z_3}-{z_4}-{z_5}-{z_7}-{z_8}-a_{1345678}+4)
  \ebrk \times
  \biggl(\Gamma(-a-3 {\ep}-{z_6}+6)
  \Gamma(-2 {z_1}-{z_2}-2 {z_3}+{z_4}-2 {z_6}+a_{78})
  \Gamma(-2 {\ep}-2 {z_6}-a_{1234}+4)
  \ebrk \times
  \Gamma(-2 {\ep}+{z_1}+{z_2}+{z_3}-{z_4}-{z_7}-a_{135678}+4)
  \ebrk \times
  \Gamma(-2 {\ep}+{z_1}+{z_2}+{z_3}-{z_4}-{z_7}-a_{245678}+4)\biggr)^{-1},
\end{eqnarray}
where $a=\sum_{i=1}^8 a_i$, $a_S = \sum_{i \in S} a_i$ with $S$ a subset of
${1,...,8}$,  and $a_i$ with $i=1,...,7$ are the powers of the denominators
according to the labeling given in Fig.~\ref{fig:npnum}, whereas $a_8$ is
the power of the additional denominator $1/k^2$. Even though Eq.~(\ref{rep2}) has
two more integrations than Eq.~(\ref{rep1}), the presence of the factor
$1/\Gamma(a_8)$ makes it necessary to perform first an analytic continuation in
$a_8$ to a negative integer value, which effectively reduces the number of
integration variables back to six. In fact, the above representation can be
used to compute any of the master integrals from the full set of non-planars
of Fig.~\ref{fig:ints}.

To complete our exposition of the direct calculation of the
amplitude, we have to note that the renormalization of the bare
amplitude requires the on-shell wave function renormalization
constant of the gluon, which is non-vanishing due to the presence of
heavy-quark loops. We give it here as an expression exact in
$d$-dimensions,
\begin{eqnarray}
  \label{eq:Z3}
  Z_3 &=& 1
  + {a^b_s} {n_h} {T_F} \left(-\frac{2}{3{\ep}}\right)
  +\left({a^b_s}\right)^2 {n_h} {T_F} \left({n_h} {T_F} \left(\frac{4}{9 {\ep}^2}\right)
    \brk +{C_F} \left(\frac{4
          {\ep}^3-7 {\ep}-1}{{\ep} \left(4
            {\ep}^3-8
            {\ep}^2-{\ep}+2\right)}\right)+{C_A}
      \left(\frac{-4 {\ep}^5+15
          {\ep}^3+{\ep}^2-11 {\ep}-3}{2 {\ep}^2
          \left(4 {\ep}^4-4 {\ep}^3-13 {\ep}^2+7
            {\ep}+6\right)}\right)\right),
\end{eqnarray}
where $a^b_s$ is defined by the bare coupling constant $\alpha_s^b$
and with $S_\epsilon=(4 \pi)^\ep \exp(-\ep \gamma_{\rm E})$, and
\begin{equation}
  \label{eq:ab-shortdef}
  a^b_s = \frac{\alpha_s^b}{2\pi}\left(\frac{\mu^2}{m^2}\right)^{\ep}
  e^{\ep \gamma_E} \Gamma(1+\ep)\;S_\ep
  \, .
\end{equation}
In terms of the renormalized coupling $\alpha_s$ and expanding in
powers of $\epsilon$ through sufficient terms, the result for $Z_3$ reads
\begin{eqnarray}
  \label{eq:Z3ren}
  Z_3 &=& 1
+ \left(\frac{\alpha_s}{2\pi}\right) \nh \tf
\biggl\{
  - {2 \over 3 \ep}
  - {2 \over 3} \log \left( \frac{\mu^2}{m^2} \right)
  - {1 \over 3} \ep \log^2 \left( \frac{\mu^2}{m^2} \right)
  - {\pi^2 \over 18} \ep
  - {1 \over 9} \ep^2 \log^3 \left( \frac{\mu^2}{m^2} \right)
\nonumber\\
& &
  - {\pi^2 \over 18} \ep^2 \log \left( \frac{\mu^2}{m^2} \right)
  + {2 \over 9} \ep^2 \z3
\biggr\}
+ \left(\frac{\alpha_s}{2\pi}\right)^2 \nh \tf
\biggl\{
  \nh \tf \biggl[
      {4 \over 9 \ep} \log \left( \frac{\mu^2}{m^2} \right)
    + {2 \over 3} \log^2 \left( \frac{\mu^2}{m^2} \right)
    + {\pi^2 \over 27}
  \biggr]
\nonumber\\
& &
+ \nl \tf \biggl[
    - {4 \over 9 \ep^2}
    - {4 \over 9 \ep} \log \left( \frac{\mu^2}{m^2} \right)
    - {2 \over 9} \log^2 \left( \frac{\mu^2}{m^2} \right)
    - {\pi^2 \over 27}
  \biggr]
+ \cf \biggl[
    - {1 \over 2 \ep}
    - \log \left( \frac{\mu^2}{m^2} \right)
    - {15 \over 4}
  \biggr]
\nonumber\\
& &
+ \ca \biggl[
      {35 \over 36 \ep^2}
    + {13 \over 18 \ep} \log \left( \frac{\mu^2}{m^2} \right)
    - {5 \over 8 \ep}
    - {5 \over 4} \log \left( \frac{\mu^2}{m^2} \right)
    + {1 \over 9} \log^2 \left( \frac{\mu^2}{m^2} \right)
    + {13 \over 48}
    + {13 \pi^2 \over 216}
  \biggr]
\biggr\}\, .
\end{eqnarray}
The corresponding expression for the wave function renormalization
constant $Z_2$ of a light quark state has been given in
Ref.~\cite{Czakon:2007ej}.

Our result for $Z_3$ in Eqs.~(\ref{eq:Z3}) and (\ref{eq:Z3ren})
coincides to first order in $\as$ with $Z^{(1)}_{[g]}$ from
Eq.~(\ref{eq:Zg}). At second order in $\as$ all known terms (i.e.
those quadratic in the number of flavors) in the two constants are
also a complete match (also in $d$-dimensions). In our direct
evaluation of $Z_3$ in Eq.~(\ref{eq:Z3}) we observe gauge
independence through two loops (within the class of covariant gauges
employed in the calculation), which is consistent 
with the arguments given in Ref.~\cite{Mitov:2006xs} in favor of the
identification of the constant $Z^{(m\vert0)}_{[g]}$ with $Z_3$
evaluated in a physical gauge.

%
%
\section{Results}
\label{sec:results}

We are now ready to present our result for $gg \to Q {\bar Q}$ scattering
for the interference of the two-loop and Born amplitude,
\begin{eqnarray}
  \label{eq:ReM0x2}
{\lefteqn{
  2 {\rm Re}\, \langle {\cal M}^{(0)} | {\cal M}^{(2)} \rangle =
(N^2-1) \biggl(
N^3 A + N B  + {1 \over N} C + {1 \over N^3} D
+ N^2 \nl E_l + N^2 \nh E_h
}}
\\
& &
+ \nl F_l + \nh F_h
+ {\nl \over N^2} G_l + {\nh \over N^2} G_h
+ N \nl^2 H_l + N \nl \nh\, H_{lh} + N \nh^2 H_h
+ {\nl^2 \over N} I_l + {\nl \nh \over N} I_{lh} + {\nh^2 \over N} I_h
\biggr)
\, ,
\nonumber
\end{eqnarray}
which we have ordered according to the power of the number of colors $N$
and the numbers of $\nl$ light and $\nh=1$ heavy quarks with $\nf=\nl+\nh$ total flavors.

The coefficients $A$, $E_l$, $H_l$, $H_{lh}$, $H_h$, $I_l$, $I_{lh}$,
and $I_h$ have been computed with both of our methods.
We have found agreement between the direct computation of the relevant Feynman diagrams
in the small-mass expansion and the QCD factorization approach as given by the
universal multiplicative relation~(\ref{eq:Mm-M0}).
All other terms linear in $\nh$, that includes $E_h$, $F_h$ and $G_h$ have been obtained
by means of a direct calculations of the massive loop integrals as detailed above.
The remaining coefficients $B$, $C$, $D$, $F_l$ and $G_l$ have been derived
by application of the factorization formula.

We choose $x=-t/s$  as the only dimensionless kinematic variable in the
problem and we keep the dependence on the renormalization scale, $\mu$, explicit.
We also introduce the following compact notation
\begin{equation}
\lm = \log\left( \frac{m^2}{s} \right)\, , \;\;\;\;
\ls = \log\left( \frac{s}{\mu^2} \right)\, , \;\;\;\;
\lx = \log\left( x \right)\, , \;\;\;\;
\ly = \log\left( 1-x \right)\, .
\end{equation}

The different components now read
{\small{
\begin{eqnarray}
A &=& \frac{1}{\ep^4}\left\{
-8 x^2+8 x+\frac{2}{1-x}-8+\frac{2}{x}
\right\}
+
\frac{1}{\ep^3}\left\{
\lm\left[
-8 x^2+8 x+\frac{2}{1-x}-8+\frac{2}{x}
\right]
+
\ls\left[
16 x^2-16 x-\frac{4}{1-x}
\ibrk
+16-\frac{4}{x}
\right]
-\frac{154 x^2}{3}+\frac{154 x}{3}+\ly \left(8
   x^2-\frac{4}{1-x}+4\right)+\lx \left(8 x^2-16
   x+12-\frac{4}{x}\right)+\frac{65}{6 (1-x)}-\frac{142}{3}
\brk
+\frac{65}{6 x}
\right\}
+
\frac{1}{\ep^2}\left\{
\lm^2\left[
2 x^2-2 x-\frac{1}{2 (1-x)}+2-\frac{1}{2 x}
\right]
+
\lm \ls\left[
16 x^2-16 x-\frac{4}{1-x}+16-\frac{4}{x}
\right]
\brk
+
\ls^2\left[
-16 x^2+16 x+\frac{4}{1-x}-16+\frac{4}{x}
\right]
+
\lm\left[
-22 x^2+22 x+\ly \left(4 x^2-\frac{2}{1-x}+2\right)
\ibrk
+\lx \left(4
   x^2-8 x+6-\frac{2}{x}\right)+\frac{7}{2 (1-x)}-18+\frac{7}{2 x}
\right]
+
\ls\left[
44 x^2-44 x+\ly \left(-16 x^2+\frac{8}{1-x}-8\right)
\ibrk
+\lx \left(-16
   x^2+32 x-24+\frac{8}{x}\right)-\frac{7}{1-x}+36-\frac{7}{x}
\right]
+
\left(-4 x^2+12 x-9+\frac{3}{x}\right) \lx^2
\brk
+\left(18 x^2-41
   x+31-\frac{8}{x}\right) \lx-\frac{104 x^2}{9}+\frac{104
   x}{9}+\ly^2 \left(-4 x^2-4 x+\frac{3}{1-x}-1\right)
\brk
+\pi ^2 \left(8
   x^2-8 x-\frac{2}{1-x}+8-\frac{2}{x}\right)+\ly \left(18 x^2+5
   x-\frac{8}{1-x}+8\right)-\frac{161}{18 (1-x)}-\frac{161}{18
   x}+\frac{100}{9}
\right\}
\ebrk
+
\frac{1}{\ep}\left\{
\lm^3\left[
\frac{2 x^2}{3}-\frac{2 x}{3}-\frac{1}{6 (1-x)}+\frac{2}{3}-\frac{1}{6 x}
\right]
+
\lm^2 \ls\left[
-4 x^2+4 x+\frac{1}{1-x}-4+\frac{1}{x}
\right]
+
\lm \ls^2\left[
-16 x^2
\ibrk
+16 x+\frac{4}{1-x}-16+\frac{4}{x}
\right]
+
\ls^3\left[
\frac{32 x^2}{3}-\frac{32 x}{3}-\frac{8}{3 (1-x)}+\frac{32}{3}-\frac{8}{3 x}
\right]
+
\lm^2\left[
\frac{22 x^2}{3}-\frac{22 x}{3}
\ibrk
+\ly \left(-2
   x^2+\frac{1}{1-x}-1\right)+\lx \left(-2 x^2+4
   x-3+\frac{1}{x}\right)-\frac{4}{3 (1-x)}+\frac{19}{3}-\frac{4}{3 x}
\right]
+
\lm \ls\left[
\frac{44 x^2}{3}-\frac{44 x}{3}
\ibrk
+\ly \left(-8
   x^2+\frac{4}{1-x}-4\right)+\lx \left(-8 x^2+16
   x-12+\frac{4}{x}\right)+\frac{1}{3 (1-x)}+\frac{20}{3}+\frac{1}{3 x}
\right]
+
\ls^2\left[
-\frac{44 x^2}{3}
\ibrk
+\frac{44 x}{3}+\ly \left(16
   x^2-\frac{8}{1-x}+8\right)+\lx \left(16 x^2-32
   x+24-\frac{8}{x}\right)-\frac{1}{3 (1-x)}-\frac{20}{3}-\frac{1}{3 x}
\right]
\brk
+
\lm\left[
\left(2 x-\frac{3}{2}+\frac{1}{2 x}\right) \lx^2+\left(-2 x^2+\frac{3
   x}{2}-1+\frac{3}{2 x}\right) \lx+\frac{140 x^2}{9}+\ly^2
   \left(-2 x+\frac{1}{2 (1-x)}+\frac{1}{2}\right)
\ibrk
-\frac{140 x}{9}+\ly
   \left(-2 x^2+\frac{5 x}{2}+\frac{3}{2 (1-x)}-\frac{3}{2}\right)+\pi ^2
   \left(2 x^2-2 x-\frac{1}{2 (1-x)}+2-\frac{1}{2 x}\right)-\frac{80}{9
   (1-x)}-\frac{80}{9 x}
\ibrk
+\frac{451}{18}
\right]
+
\ls\left[
\left(8 x^2-24 x+18-\frac{6}{x}\right) \lx^2+\left(-\frac{20
   x^2}{3}+\frac{70 x}{3}-18+\frac{4}{3 x}\right) \lx-\frac{496
   x^2}{9}+\frac{496 x}{9}
\ibrk
+\pi ^2 \left(-16 x^2+16
   x+\frac{4}{1-x}-16+\frac{4}{x}\right)+\ly \left(-\frac{20 x^2}{3}-10
   x+\frac{4}{3 (1-x)}-\frac{4}{3}\right)
\ibrk
+\ly^2 \left(8 x^2+8
   x-\frac{6}{1-x}+2\right)+\frac{205}{9 (1-x)}+\frac{205}{9 x}-\frac{640}{9}
\right]
+
\left(-\frac{4 x^2}{3}-4 x+3-\frac{1}{x}\right) \lx^3
\brk
+\left(4
   x^2+\frac{29 x}{6}+\ly \left(4 x^2-4
   x+3-\frac{1}{x}\right)-10+\frac{5}{6 x}\right) \lx^2+\li2
   \left(8 x^2-8 x+6-\frac{2}{x}\right) \lx
\brk
+\left(-\frac{332
   x^2}{9}+\frac{514 x}{9}+\pi ^2 \left(-\frac{34 x^2}{3}+\frac{44
   x}{3}-11+\frac{11}{3 x}\right)-\frac{116}{3}+\frac{166}{9 x}\right)
   \lx+\frac{2584 x^2}{27}-\frac{2584 x}{27}
\brk
+\li3 \left(-8 x^2+8
   x-6+\frac{2}{x}\right)+\ly^3 \left(-\frac{4 x^2}{3}+\frac{20
   x}{3}-\frac{1}{1-x}-\frac{7}{3}\right)+\ly^2 \left(4 x^2-\frac{77
   x}{6}+\frac{5}{6 (1-x)}-\frac{7}{6}\right)
\brk
+\s12 \left(8 x^2-8
   x-\frac{2}{1-x}+6\right)+\pi ^2 \left(19 x^2-19 x+\frac{1}{4
   (1-x)}+4+\frac{1}{4 x}\right)+\ly \left(-\frac{332 x^2}{9}+\frac{50
   x}{3}
\ibrk
+\pi ^2 \left(-\frac{34 x^2}{3}+8 x+\frac{11}{3
   (1-x)}-\frac{23}{3}\right)+\frac{166}{9
   (1-x)}-\frac{166}{9}\right)+\left(\frac{184 x^2}{3}-\frac{184
   x}{3}-\frac{40}{3 (1-x)}+\frac{178}{3}
\ibrk
-\frac{46}{3 x}\right)
   \z3-\frac{625}{27 (1-x)}-\frac{625}{27 x}+\frac{2416}{27}
\right\}
+
\lm^4\left[
-\frac{5 x^2}{6}+\frac{5 x}{6}+\frac{5}{24 (1-x)}-\frac{5}{6}+\frac{5}{24 x}
\right]
\ebrk
+
\lm^3 \ls\left[
-\frac{4 x^2}{3}+\frac{4 x}{3}+\frac{1}{3 (1-x)}-\frac{4}{3}+\frac{1}{3 x}
\right]
+
\lm^2 \ls^2\left[
4 x^2-4 x-\frac{1}{1-x}+4-\frac{1}{x}
\right]
\ebrk
+
\lm \ls^3\left[
\frac{32 x^2}{3}-\frac{32 x}{3}-\frac{8}{3 (1-x)}+\frac{32}{3}-\frac{8}{3 x}
\right]
+
\ls^4\left[
-\frac{16 x^2}{3}+\frac{16 x}{3}+\frac{4}{3 (1-x)}-\frac{16}{3}+\frac{4}{3 x}
\right]
\ebrk
+
\lm^3\left[
\ly \left(\frac{2 x^2}{3}-\frac{1}{3 (1-x)}+\frac{1}{3}\right)+\lx
   \left(\frac{2 x^2}{3}-\frac{4 x}{3}+1-\frac{1}{3 x}\right)+\frac{1}{6
   (1-x)}+\frac{1}{6 x}-\frac{1}{3}
\right]
\ebrk
+
\lm^2 \ls\left[
\ly \left(4 x^2-\frac{2}{1-x}+2\right)+\lx \left(4 x^2-8
   x+6-\frac{2}{x}\right)-\frac{1}{1-x}-\frac{1}{x}+2
\right]
\ebrk
+
\lm \ls^2\left[
\ly \left(8 x^2-\frac{4}{1-x}+4\right)+\lx \left(8 x^2-16
   x+12-\frac{4}{x}\right)-\frac{4}{1-x}-\frac{4}{x}+8
\right]
\ebrk
+
\ls^3\left[
\ly \left(-\frac{32 x^2}{3}+\frac{16}{3
   (1-x)}-\frac{16}{3}\right)+\lx \left(-\frac{32 x^2}{3}+\frac{64
   x}{3}-16+\frac{16}{3 x}\right)+\frac{8}{3 (1-x)}+\frac{8}{3 x}-\frac{16}{3}
\right]
\ebrk
+
\lm^2\left[
\left(-x+\frac{3}{4}-\frac{1}{4 x}\right) \lx^2+\left(x^2-\frac{3
   x}{4}+\frac{1}{2}-\frac{3}{4 x}\right) \lx-\frac{391
   x^2}{18}+\frac{391 x}{18}+\ly^2 \left(x-\frac{1}{4
   (1-x)}-\frac{1}{4}\right)
\brk
+\pi ^2 \left(-x^2+x+\frac{1}{4
   (1-x)}-1+\frac{1}{4 x}\right)+\ly \left(x^2-\frac{5 x}{4}-\frac{3}{4
   (1-x)}+\frac{3}{4}\right)+\frac{505}{72 (1-x)}+\frac{505}{72
   x}-\frac{887}{36}
\right]
\ebrk
+
\lm \ls\left[
\left(-4 x+3-\frac{1}{x}\right) \lx^2+\left(4 x^2-3
   x+2-\frac{3}{x}\right) \lx-\frac{148 x^2}{9}+\frac{148
   x}{9}+\ly^2 \left(4 x-\frac{1}{1-x}-1\right)
\brk
+\pi ^2 \left(-4 x^2+4
   x+\frac{1}{1-x}-4+\frac{1}{x}\right)+\ly \left(4 x^2-5
   x-\frac{3}{1-x}+3\right)+\frac{61}{9 (1-x)}+\frac{61}{9 x}-\frac{187}{9}
\right]
\ebrk
+
\ls^2\left[
\left(-8 x^2+24 x-18+\frac{6}{x}\right) \lx^2+\left(-8 x^2+6
   x-4+\frac{6}{x}\right) \lx+\frac{364 x^2}{9}-\frac{364
   x}{9}
\brk
+\ly^2 \left(-8 x^2-8 x+\frac{6}{1-x}-2\right)+\ly
   \left(-8 x^2+10 x+\frac{6}{1-x}-6\right)+\pi ^2 \left(16 x^2-16
   x-\frac{4}{1-x}+16-\frac{4}{x}\right)
\brk
-\frac{106}{9 (1-x)}-\frac{106}{9
   x}+\frac{376}{9}
\right]
+
\lm\left[
-\frac{2}{3} x^2 \lx^3+\left(\frac{x}{4}+\ly \left(2 x^2-2
   x+\frac{3}{2}-\frac{1}{2 x}\right)-\frac{9}{4}\right)
   \lx^2
\brk
+\li2 \left(4 x^2-4 x+3-\frac{1}{x}\right)
   \lx+\left(6 x^2-\frac{61 x}{4}+\pi ^2 \left(-2 x^2+2
   x-\frac{3}{2}+\frac{1}{2 x}\right)+\frac{29}{2}-\frac{21}{4 x}\right)
   \lx-\frac{1165 x^2}{54}
\brk
+\ly^2
   \left(-\frac{x}{4}-2\right)+\frac{1165 x}{54}+\li3 \left(-4 x^2+4
   x-3+\frac{1}{x}\right)+\ly^3 \left(-\frac{2 x^2}{3}+\frac{4
   x}{3}-\frac{2}{3}\right)
\brk
+\s12 \left(4 x^2-4
   x-\frac{1}{1-x}+3\right)+\pi ^2 \left(\frac{17 x^2}{3}-\frac{17
   x}{3}-\frac{1}{6 (1-x)}+\frac{23}{12}-\frac{1}{6 x}\right)+\ly
   \left(6 x^2+\frac{13 x}{4}
\ibrk
+\pi ^2 \left(-2 x^2+2 x+\frac{1}{2
   (1-x)}-\frac{3}{2}\right)-\frac{21}{4
   (1-x)}+\frac{21}{4}\right)+\left(-\frac{8 x^2}{3}+\frac{8 x}{3}+\frac{5}{3
   (1-x)}-\frac{11}{3}+\frac{2}{3 x}\right) \z3
\brk
+\frac{2203}{216
   (1-x)}+\frac{2203}{216 x}-\frac{3521}{108}
\right]
+
\ls\left[
\left(\frac{8 x^2}{3}+8 x-6+\frac{2}{x}\right) \lx^3+\left(-8 x^2+5
   x
\ibrk
+\ly \left(-8 x^2+8 x-6+\frac{2}{x}\right)+9+\frac{2}{x}\right)
   \lx^2+\li2 \left(-16 x^2+16 x-12+\frac{4}{x}\right)
   \lx+\left(\frac{400 x^2}{9}-\frac{665 x}{9}
\ibrk
+\pi ^2 \left(\frac{68
   x^2}{3}-\frac{88 x}{3}+22-\frac{22}{3 x}\right)+48-\frac{167}{9 x}\right)
   \lx-\frac{1040 x^2}{27}+\frac{1040 x}{27}
\brk
+\s12 \left(-16 x^2+16
   x+\frac{4}{1-x}-12\right)+\pi ^2 \left(-\frac{122 x^2}{9}+\frac{122
   x}{9}-\frac{119}{18 (1-x)}+\frac{148}{9}-\frac{119}{18
   x}\right)
\brk
+\ly^2 \left(-8 x^2+11 x+\frac{2}{1-x}+6\right)+\ly^3
   \left(\frac{8 x^2}{3}-\frac{40
   x}{3}+\frac{2}{1-x}+\frac{14}{3}\right)+\li3 \left(16 x^2-16
   x+12-\frac{4}{x}\right)
\brk
+\ly \left(\frac{400 x^2}{9}-15 x+\pi ^2
   \left(\frac{68 x^2}{3}-16 x-\frac{22}{3
   (1-x)}+\frac{46}{3}\right)-\frac{167}{9
   (1-x)}+\frac{167}{9}\right)
\brk
+\left(-\frac{368 x^2}{3}+\frac{368
   x}{3}+\frac{80}{3 (1-x)}-\frac{356}{3}+\frac{92}{3 x}\right)
   \z3-\frac{211}{27 (1-x)}-\frac{211}{27 x}+\frac{55}{27}
\right]
\ebrk
+
\left(\frac{13 x^2}{6}+\frac{x}{6}-\frac{1}{8}+\frac{1}{24 x}\right)
   \lx^4+\left(\frac{2 x^2}{3}-\frac{17 x}{12}+\ly \left(-\frac{22
   x^2}{3}+8 x-6+\frac{2}{x}\right)+\frac{26}{3}-\frac{3}{4 x}\right)
   \lx^3
\ebrk
+\left(\left(2 x^2+2 x-\frac{15}{2}+\frac{9}{2 x}\right)
   \ly^2+\left(-2 x^2-\frac{59 x}{12}+1+\frac{31}{12 x}\right)
   \ly-\frac{265 x}{72}+\pi ^2 \left(12 x^2-\frac{38
   x}{3}+\frac{19}{2}-\frac{19}{6 x}\right)
\brk
-\frac{5}{1-x}-\frac{91}{72
   x}-\frac{17}{12}\right) \lx^2+\left(-\frac{262 x^2}{27}+\frac{3379
   x}{216}+\ly \pi ^2 \left(-\frac{20 x^2}{3}+\frac{8
   x}{3}+2-\frac{2}{x}\right)
\brk
+\pi ^2 \left(-\frac{17 x^2}{9}+4
   x+10-\frac{16}{3 x}\right)+\left(-\frac{172 x^2}{3}+\frac{272
   x}{3}-68+\frac{68}{3 x}\right) \z3-\frac{97}{18}-\frac{2279}{216
   x}\right) \lx-\frac{1246 x^2}{81}
\ebrk
+\s22 \left(-32
   x+\frac{18}{1-x}+16-\frac{18}{x}\right)+\frac{1246 x}{81}+\pi ^2
   \left(-\frac{422 x^2}{27}+\frac{422 x}{27}+\frac{589}{54
   (1-x)}-\frac{3581}{108}+\frac{589}{54 x}\right)
\ebrk
+\s13 \left(-4 x^2-4
   x+\frac{3}{1-x}-1\right)+\pi ^4 \left(\frac{17 x^2}{30}+\frac{7
   x}{30}-\frac{173}{360 (1-x)}+\frac{121}{180}-\frac{23}{360
   x}\right)
\ebrk
+\ly^3 \left(\frac{2 x^2}{3}+\frac{x}{12}-\frac{3}{4
   (1-x)}+\frac{95}{12}\right)+\ly^4 \left(\frac{13 x^2}{6}-\frac{9
   x}{2}+\frac{1}{24 (1-x)}+\frac{53}{24}\right)+\li4 \left(4 x^2-12
   x+9-\frac{3}{x}\right)
\ebrk
+\s12 \left(-4 x^2+\frac{107 x}{6}+\ly
   \left(-16 x^2+12 x+\frac{5}{1-x}-11\right)+\lx \left(8 x^2+8
   x-30+\frac{18}{x}\right)+\frac{31}{6 (1-x)}
\brk
-\frac{71}{6}\right)+\ly^2
   \left(\frac{265 x}{72}+\pi ^2 \left(12 x^2-\frac{34 x}{3}-\frac{19}{6
   (1-x)}+\frac{53}{6}\right)-\frac{91}{72
   (1-x)}-\frac{367}{72}-\frac{5}{x}\right)
\ebrk
+\li3 \left(4 x^2+\frac{59
   x}{6}+\ly \left(-8 x^2-8 x+30-\frac{18}{x}\right)+\lx \left(12
   x^2-8 x+6-\frac{2}{x}\right)-2-\frac{31}{6 x}\right)
\ebrk
+\li2
   \left(\left(-14 x^2+14 x-\frac{21}{2}+\frac{7}{2 x}\right)
   \lx^2+\left(-4 x^2-\frac{59 x}{6}+\ly \left(8 x^2+8
   x-30+\frac{18}{x}\right)+2+\frac{31}{6 x}\right) \lx
\brk
+\pi ^2 \left(-8
   x+\frac{2}{1-x}+4-\frac{2}{x}\right)\right)+\left(52 x^2-\frac{395
   x}{6}+\frac{23}{6 (1-x)}+\frac{89}{6}+\frac{9}{x}\right)
   \z3+\ly \left(-\frac{262 x^2}{27}+\frac{271 x}{72}
\brk
+\pi ^2
   \left(-\frac{17 x^2}{9}-\frac{2 x}{9}-\frac{16}{3
   (1-x)}+\frac{109}{9}\right)+\left(-\frac{100 x^2}{3}+20 x+\frac{53}{3
   (1-x)}-\frac{161}{3}+\frac{18}{x}\right) \z3
\brk
-\frac{2279}{216
   (1-x)}+\frac{119}{216}\right)+\frac{16313}{648 (1-x)}+\frac{16313}{648
   x}-\frac{39113}{648},
\end{eqnarray}

\begin{eqnarray}
B &=& \frac{1}{\ep^4}\left\{
-\frac{2}{x}+4-\frac{2}{1-x}
\right\}
+
\frac{1}{\ep^3}\left\{
\lm\left[
8 x^2-8 x-\frac{4}{1-x}+12-\frac{4}{x}
\right]
+
\ls\left[
\frac{4}{x}-8+\frac{4}{1-x}
\right]
+
8 x^2-8 x
\brk
+\lx \left(\frac{4}{x}-8+\frac{4}{1-x}\right)+\ly
   \left(\frac{4}{x}-8+\frac{4}{1-x}\right)-\frac{77}{6
   (1-x)}+\frac{101}{3}-\frac{77}{6 x}
\right\}
+
\frac{1}{\ep^2}\left\{
\lm^2\left[
\frac{1}{2 x}-1+\frac{1}{2 (1-x)}
\right]
\brk
+
\lm \ls\left[
-16 x^2+16 x+\frac{8}{1-x}-24+\frac{8}{x}
\right]
+
\ls^2\left[
-\frac{4}{x}+8-\frac{4}{1-x}
\right]
+
\lm\left[
26 x^2-26 x
\ibrk
+\ly \left(-4
   x^2+\frac{4}{1-x}-6+\frac{2}{x}\right)+\lx \left(-4 x^2+8
   x+\frac{2}{1-x}-10+\frac{4}{x}\right)-\frac{8}{1-x}+33-\frac{8}{x}
\right]
\brk
+
\ls\left[
-16 x^2+16 x+\lx \left(-\frac{8}{x}+16-\frac{8}{1-x}\right)+\ly
   \left(-\frac{8}{x}+16-\frac{8}{1-x}\right)+\frac{11}{1-x}-38+\frac{11}{x}
\right]
\brk
+
\left(-8 x^2+x+1-\frac{2}{x}\right) \lx^2+\left(-4 x^2+9 x+\ly
   \left(16 x^2-16
   x-\frac{8}{1-x}+24-\frac{8}{x}\right)+\frac{5}{1-x}-33+\frac{13}{x}\right)
   \lx
\brk
+34 x^2-34 x+\pi ^2 \left(-\frac{28 x^2}{3}+\frac{28
   x}{3}+\frac{1}{3 (1-x)}-\frac{16}{3}+\frac{1}{3 x}\right)+\ly^2
   \left(-8 x^2+15 x-\frac{2}{1-x}-6\right)
\brk
+\ly \left(-4
   x^2-x+\frac{13}{1-x}-28+\frac{5}{x}\right)+\frac{31}{9 (1-x)}+\frac{31}{9
   x}+\frac{313}{9}
\right\}
+
\frac{1}{\ep}\left\{
\lm^3\left[
-\frac{8 x^2}{3}+\frac{8 x}{3}+\frac{5}{6 (1-x)}-3
\ibrk
+\frac{5}{6 x}
\right]
+
\lm^2 \ls\left[
-\frac{1}{x}+2-\frac{1}{1-x}
\right]
+
\lm \ls^2\left[
16 x^2-16 x-\frac{8}{1-x}+24-\frac{8}{x}
\right]
+
\ls^3\left[
\frac{8}{3 x}-\frac{16}{3}+\frac{8}{3 (1-x)}
\right]
\brk
+
\lm^2\left[
-\frac{28 x^2}{3}+\frac{28 x}{3}+\ly \left(2
   x^2-\frac{2}{1-x}+3-\frac{1}{x}\right)+\lx \left(2 x^2-4
   x-\frac{1}{1-x}+5-\frac{2}{x}\right)+\frac{11}{3 (1-x)}-13
\ibrk
+\frac{11}{3 x}
\right]
+
\lm \ls\left[
-\frac{68 x^2}{3}+\frac{68 x}{3}+\ly \left(8
   x^2-\frac{8}{1-x}+12-\frac{4}{x}\right)+\lx \left(8 x^2-16
   x-\frac{4}{1-x}+20-\frac{8}{x}\right)
\ibrk
+\frac{4}{3 (1-x)}-22+\frac{4}{3 x}
\right]
+
\ls^2\left[
16 x^2-16 x+\lx \left(\frac{8}{x}-16+\frac{8}{1-x}\right)+\ly
   \left(\frac{8}{x}-16+\frac{8}{1-x}\right)-\frac{11}{3
   (1-x)}
\ibrk
+\frac{70}{3}-\frac{11}{3 x}
\right]
+
\lm\left[
\left(-\frac{3 x}{2}-\frac{1}{1-x}+2-\frac{1}{2 x}\right) \lx^2+\left(2
   x^2-x+\ly
   \left(-\frac{2}{x}+4-\frac{2}{1-x}\right)-\frac{3}{1-x}-\frac{3}{2}
\iibrk
-\frac{3
   }{2 x}\right) \lx-\frac{86 x^2}{9}+\frac{86 x}{9}+\ly^2
   \left(\frac{3 x}{2}-\frac{1}{2 (1-x)}+\frac{1}{2}-\frac{1}{x}\right)+\pi ^2
   \left(-\frac{14 x^2}{3}+\frac{14 x}{3}+\frac{5}{3
   (1-x)}-\frac{17}{3}+\frac{5}{3 x}\right)
\ibrk
+\ly \left(2 x^2-3
   x-\frac{3}{2 (1-x)}-\frac{1}{2}-\frac{3}{x}\right)+\frac{160}{9
   (1-x)}+\frac{160}{9 x}-\frac{88}{3}
\right]
+
\ls\left[
\left(16 x^2-2 x-2+\frac{4}{x}\right) \lx^2
\ibrk
+\left(8 x^2-18 x+\ly
   \left(-32 x^2+32 x+\frac{16}{1-x}-48+\frac{16}{x}\right)+\frac{14}{3
   (1-x)}+\frac{110}{3}-\frac{34}{3 x}\right) \lx-\frac{116
   x^2}{3}+\frac{116 x}{3}
\ibrk
+\ly \left(8 x^2+2 x-\frac{34}{3
   (1-x)}+\frac{80}{3}+\frac{14}{3 x}\right)+\ly^2 \left(16 x^2-30
   x+\frac{4}{1-x}+12\right)
\ibrk
+\pi ^2 \left(\frac{56 x^2}{3}-\frac{56
   x}{3}-\frac{2}{3 (1-x)}+\frac{32}{3}-\frac{2}{3 x}\right)-\frac{172}{9
   (1-x)}-\frac{172}{9 x}-\frac{10}{9}
\right]
+
\left(-\frac{23 x}{3}+\frac{8}{3 (1-x)}-\frac{1}{3}-\frac{2}{3 x}\right)
   \lx^3
\brk
+\left(8 x^2-\frac{19 x}{6}+\ly \left(23
   x-\frac{4}{1-x}-14+\frac{5}{x}\right)-\frac{20}{3
   (1-x)}+\frac{82}{3}-\frac{11}{2 x}\right) \lx^2+\li2 \left(2
   x-\frac{8}{1-x}+10
\ibrk
-\frac{4}{x}\right) \lx+\left(\left(-22
   x+\frac{7}{1-x}+3\right) \ly^2+\left(-16 x^2+16 x+\frac{11}{3
   (1-x)}-\frac{46}{3}+\frac{11}{3 x}\right) \ly-4 x^2+\frac{95
   x}{6}
\ibrk
+\pi ^2 \left(\frac{20 x^2}{3}+\frac{14 x}{3}-\frac{11}{3
   (1-x)}+\frac{31}{3}\right)-\frac{238}{9 (1-x)}-\frac{179}{18
   x}-\frac{10}{9}\right) \lx+\frac{184 x^2}{9}
\brk
+\s12 \left(-2
   x-\frac{4}{1-x}+12-\frac{8}{x}\right)+\li3 \left(-2
   x+\frac{8}{1-x}-10+\frac{4}{x}\right)-\frac{184 x}{9}
\brk
+\ly^3
   \left(\frac{23 x}{3}-\frac{2}{3 (1-x)}-8+\frac{8}{3 x}\right)+\pi ^2
   \left(-\frac{226 x^2}{9}+\frac{226 x}{9}+\frac{259}{36
   (1-x)}-\frac{121}{9}+\frac{259}{36 x}\right)
\brk
+\ly^2 \left(8
   x^2-\frac{77 x}{6}-\frac{11}{2 (1-x)}+\frac{193}{6}-\frac{20}{3
   x}\right)+\ly \left(-4 x^2-\frac{47 x}{6}+\pi ^2 \left(\frac{20
   x^2}{3}-18 x+\frac{65}{3}-\frac{11}{3 x}\right)
\ibrk
-\frac{179}{18
   (1-x)}+\frac{193}{18}-\frac{238}{9 x}\right)+\left(-20 x^2+22 x+\frac{55}{3
   (1-x)}-\frac{158}{3}+\frac{67}{3 x}\right) \z3+\frac{820}{27
   (1-x)}+\frac{820}{27 x}-\frac{2569}{54}
\right\}
\ebrk
+
\lm^4\left[
2 x^2-2 x-\frac{17}{24 (1-x)}+\frac{29}{12}-\frac{17}{24 x}
\right]
+
\lm^3 \ls\left[
\frac{16 x^2}{3}-\frac{16 x}{3}-\frac{5}{3 (1-x)}+6-\frac{5}{3 x}
\right]
\ebrk
+
\lm^2 \ls^2\left[
\frac{1}{x}-2+\frac{1}{1-x}
\right]
+
\lm \ls^3\left[
-\frac{32 x^2}{3}+\frac{32 x}{3}+\frac{16}{3 (1-x)}-16+\frac{16}{3 x}
\right]
+
\ls^4\left[
-\frac{4}{3 x}+\frac{8}{3}-\frac{4}{3 (1-x)}
\right]
\ebrk
+
\lm^3\left[
\frac{2 x^2}{3}-\frac{2 x}{3}+\ly \left(-\frac{2 x^2}{3}+\frac{2}{3
   (1-x)}-1+\frac{1}{3 x}\right)+\lx \left(-\frac{2 x^2}{3}+\frac{4
   x}{3}+\frac{1}{3 (1-x)}-\frac{5}{3}+\frac{2}{3
   x}\right)-\frac{1}{1-x}
\brk
+2-\frac{1}{x}
\right]
+
\lm^2 \ls\left[
4 x^2-4 x+\ly \left(-4 x^2+\frac{4}{1-x}-6+\frac{2}{x}\right)+\lx
   \left(-4 x^2+8 x+\frac{2}{1-x}-10+\frac{4}{x}\right)+4
\right]
\ebrk
+
\lm \ls^2\left[
8 x^2-8 x+\ly \left(-8 x^2+\frac{8}{1-x}-12+\frac{4}{x}\right)+\lx
   \left(-8 x^2+16
   x+\frac{4}{1-x}-20+\frac{8}{x}\right)+\frac{6}{1-x}+\frac{6}{x}
\right]
\ebrk
+
\ls^3\left[
-\frac{32 x^2}{3}+\frac{32 x}{3}+\lx \left(-\frac{16}{3
   x}+\frac{32}{3}-\frac{16}{3 (1-x)}\right)+\ly \left(-\frac{16}{3
   x}+\frac{32}{3}-\frac{16}{3 (1-x)}\right)-\frac{32}{3}
\right]
\ebrk
+
\lm^2\left[
\left(\frac{3 x}{4}+\frac{1}{2 (1-x)}-1+\frac{1}{4 x}\right)
   \lx^2+\left(-x^2+\frac{x}{2}+\ly
   \left(\frac{1}{x}-2+\frac{1}{1-x}\right)+\frac{3}{2
   (1-x)}+\frac{3}{4}+\frac{3}{4 x}\right) \lx
\brk
+\frac{227
   x^2}{9}+\ly^2 \left(-\frac{3 x}{4}+\frac{1}{4
   (1-x)}-\frac{1}{4}+\frac{1}{2 x}\right)-\frac{227 x}{9}+\ly
   \left(-x^2+\frac{3 x}{2}+\frac{3}{4 (1-x)}+\frac{1}{4}+\frac{3}{2
   x}\right)
\brk
+\pi ^2 \left(\frac{8 x^2}{3}-\frac{8 x}{3}-\frac{11}{12
   (1-x)}+\frac{19}{6}-\frac{11}{12 x}\right)-\frac{1127}{72
   (1-x)}-\frac{1127}{72 x}+\frac{161}{4}
\right]
\ebrk
+
\lm \ls\left[
\left(3 x+\frac{2}{1-x}-4+\frac{1}{x}\right) \lx^2+\left(-4 x^2+2
   x+\ly
   \left(\frac{4}{x}-8+\frac{4}{1-x}\right)+\frac{6}{1-x}+3+\frac{3}{x}\right)
   \lx+\frac{40 x^2}{9}
\brk
+\ly^2 \left(-3
   x+\frac{1}{1-x}-1+\frac{2}{x}\right)-\frac{40 x}{9}+\ly \left(-4
   x^2+6 x+\frac{3}{1-x}+1+\frac{6}{x}\right)
\brk
+\pi ^2 \left(\frac{28
   x^2}{3}-\frac{28 x}{3}-\frac{10}{3 (1-x)}+\frac{34}{3}-\frac{10}{3
   x}\right)-\frac{122}{9 (1-x)}-\frac{122}{9 x}+22
\right]
+
\ls^2\left[
\left(-16 x^2+2 x+2-\frac{4}{x}\right) \lx^2
\brk
+\left(-8 x^2+18 x+\ly
   \left(32 x^2-32
   x-\frac{16}{1-x}+48-\frac{16}{x}\right)-\frac{12}{1-x}-22+\frac{4}{x}\right
   ) \lx+24 x^2-24 x
\brk
+\pi ^2 \left(-\frac{56 x^2}{3}+\frac{56
   x}{3}+\frac{2}{3 (1-x)}-\frac{32}{3}+\frac{2}{3 x}\right)+\ly^2
   \left(-16 x^2+30 x-\frac{4}{1-x}-12\right)
\brk
+\ly \left(-8 x^2-2
   x+\frac{4}{1-x}-12-\frac{12}{x}\right)+\frac{106}{9 (1-x)}+\frac{106}{9
   x}-\frac{56}{9}
\right]
+
\lm\left[
\left(\frac{2 x^2}{3}-\frac{x}{3}+\frac{1}{3 (1-x)}+\frac{1}{3}
\ibrk
-\frac{1}{3
   x}\right) \lx^3+\left(-\frac{5 x}{4}+\ly \left(-2 x^2+\frac{5
   x}{2}-\frac{1}{1-x}-1+\frac{1}{2 x}\right)+\frac{1}{2
   (1-x)}+\frac{31}{4}-\frac{3}{2 x}\right) \lx^2
\brk
+\li2 \left(-4
   x^2+5 x-\frac{4}{1-x}+2-\frac{1}{x}\right)
   \lx+\left(\left(\frac{1}{x}-2+\frac{1}{1-x}\right)
   \ly^2+\left(\frac{3}{x}-8+\frac{3}{1-x}\right) \ly-6
   x^2
\ibrk
+\frac{35 x}{2}+\pi ^2 \left(2 x^2-\frac{5 x}{2}+3-\frac{3}{2
   x}\right)+\frac{4}{1-x}+\frac{47}{4 x}-\frac{109}{4}\right)
   \lx+\frac{1028 x^2}{27}-\frac{1028 x}{27}
\brk
+\ly^2 \left(\frac{5
   x}{4}-\frac{3}{2 (1-x)}+\frac{13}{2}+\frac{1}{2 x}\right)+\pi ^2
   \left(-\frac{17 x^2}{3}+\frac{17 x}{3}-\frac{13}{12
   (1-x)}+\frac{23}{6}-\frac{13}{12 x}\right)
\brk
+\s12 \left(-4 x^2+3
   x-\frac{1}{1-x}+3-\frac{4}{x}\right)+\ly^3 \left(\frac{2
   x^2}{3}-x-\frac{1}{3 (1-x)}+\frac{2}{3}+\frac{1}{3 x}\right)
\brk
+\li3
   \left(4 x^2-5 x+\frac{4}{1-x}-2+\frac{1}{x}\right)+\ly \left(-6
   x^2-\frac{11 x}{2}+\pi ^2 \left(2 x^2-\frac{3 x}{2}-\frac{3}{2
   (1-x)}+\frac{5}{2}\right)
\ibrk
+\frac{47}{4
   (1-x)}-\frac{63}{4}+\frac{4}{x}\right)+\left(-\frac{88 x^2}{3}+\frac{91
   x}{3}+\frac{14}{3 (1-x)}-28+\frac{23}{3 x}\right) \z3-\frac{4703}{216
   (1-x)}-\frac{4703}{216 x}+\frac{1997}{36}
\right]
\ebrk
+
\ls\left[
\left(\frac{46 x}{3}-\frac{16}{3 (1-x)}+\frac{2}{3}+\frac{4}{3 x}\right)
   \lx^3+\left(-16 x^2+10 x+\ly \left(-46
   x+\frac{8}{1-x}+28-\frac{10}{x}\right)+\frac{6}{1-x}
\ibrk
-51+\frac{11}{x}\right)
   \lx^2+\li2 \left(-4 x+\frac{16}{1-x}-20+\frac{8}{x}\right)
   \lx+\left(\left(44 x-\frac{14}{1-x}-6\right) \ly^2+\left(32
   x^2-32 x
\iibrk
-\frac{22}{1-x}+60-\frac{22}{x}\right) \ly+8 x^2-28 x+\pi ^2
   \left(-\frac{40 x^2}{3}-\frac{28 x}{3}+\frac{22}{3
   (1-x)}-\frac{62}{3}\right)+\frac{212}{9 (1-x)}+\frac{113}{9
   x}-\frac{13}{9}\right) \lx
\brk
-\frac{104 x^2}{9}+\ly^3
   \left(-\frac{46 x}{3}+\frac{4}{3 (1-x)}+16-\frac{16}{3 x}\right)+\frac{104
   x}{9}+\li3 \left(4 x-\frac{16}{1-x}+20-\frac{8}{x}\right)
\brk
+\s12
   \left(4 x+\frac{8}{1-x}-24+\frac{16}{x}\right)+\ly^2 \left(-16 x^2+22
   x+\frac{11}{1-x}-57+\frac{6}{x}\right)+\pi ^2 \left(\frac{92
   x^2}{3}-\frac{92 x}{3}
\ibrk
-\frac{61}{18 (1-x)}-\frac{44}{9}-\frac{61}{18
   x}\right)+\ly \left(8 x^2+12 x+\pi ^2 \left(-\frac{40 x^2}{3}+36
   x-\frac{130}{3}+\frac{22}{3 x}\right)+\frac{113}{9
   (1-x)}-\frac{193}{9}
\ibrk
+\frac{212}{9 x}\right)+\left(40 x^2-44 x-\frac{110}{3
   (1-x)}+\frac{316}{3}-\frac{134}{3 x}\right) \z3-\frac{80}{27
   (1-x)}-\frac{80}{27 x}+\frac{802}{27}
\right]
+
\left(\frac{25 x^2}{6}+\frac{57 x}{8}
\brk
-\frac{8}{3
   (1-x)}+\frac{5}{12}+\frac{9}{8 x}\right) \lx^4+\left(\frac{2
   x^2}{3}+\frac{37 x}{2}+\ly \left(-\frac{50 x^2}{3}+\frac{5
   x}{3}+\frac{2}{1-x}-4+\frac{1}{3 x}\right)-\frac{19}{2
   (1-x)}
\brk
-\frac{121}{12}+\frac{23}{4 x}\right) \lx^3+\left(\left(8
   x^2-27 x+\frac{9}{2 (1-x)}+5+\frac{5}{x}\right) \ly^2+\left(-2
   x^2-\frac{155 x}{3}+\frac{22}{3 (1-x)}+\frac{385}{12}-\frac{11}{4 x}\right)
   \ly
\brk
+\frac{x^2}{2}-\frac{539 x}{36}+\pi ^2 \left(\frac{44
   x^2}{3}+\frac{4 x}{3}-\frac{23}{6 (1-x)}+\frac{5}{6}-\frac{4}{3
   x}\right)+\frac{1093}{36 (1-x)}-\frac{83}{8
   x}-\frac{1}{(x-1)^2}+\frac{695}{72}\right) \lx^2
\ebrk
+\left(\left(-\frac{8
   x^2}{3}+\frac{61 x}{3}-\frac{11}{3 (1-x)}-\frac{8}{3}-\frac{2}{3 x}\right)
   \ly^3+\left(\frac{395 x}{6}-\frac{17}{2
   (1-x)}-\frac{307}{6}+\frac{89}{6 x}\right) \ly^2+\left(-x^2+x
\ibrk
+\pi ^2
   \left(-\frac{88 x^2}{3}+\frac{151 x}{3}+\frac{7}{3
   (1-x)}-\frac{100}{3}+\frac{41}{3 x}\right)+\frac{194}{9
   (1-x)}-\frac{142}{9}+\frac{194}{9 x}\right) \ly-8 x^2+\frac{749
   x}{18}
\brk
+\pi ^2 \left(\frac{16 x^2}{3}+\frac{221 x}{18}+\frac{73}{18
   (1-x)}-\frac{19}{6}-\frac{101}{18 x}\right)+\left(-8 x^2-16 x-\frac{80}{3
   (1-x)}+\frac{214}{3}-\frac{92}{3 x}\right) \z3
\brk
+\frac{430}{27
   (1-x)}+\frac{839}{216 x}-\frac{3457}{216}\right) \lx+\frac{308
   x^2}{27}-\frac{308 x}{27}+\s22 \left(40
   x+\frac{78}{1-x}-20-\frac{78}{x}\right)
\ebrk
+\li4 \left(-36 x^2+37
   x-\frac{74}{1-x}+96-\frac{9}{x}\right)+\ly^3 \left(\frac{2
   x^2}{3}-\frac{119 x}{6}+\frac{23}{4 (1-x)}+\frac{109}{12}-\frac{19}{2
   x}\right)
\ebrk
+\pi ^4 \left(\frac{43 x^2}{18}-\frac{268 x}{45}+\frac{557}{180
   (1-x)}-\frac{347}{90}+\frac{7}{10 x}\right)+\ly^4 \left(\frac{25
   x^2}{6}-\frac{371 x}{24}+\frac{9}{8 (1-x)}+\frac{281}{24}-\frac{8}{3
   x}\right)
\ebrk
+\s13 \left(36 x^2-35
   x+\frac{9}{1-x}-97+\frac{74}{x}\right)+\pi ^2 \left(-\frac{1}{54} (432
   \log(2)+551) x^2
\brk
+\frac{1}{54} (432 \log(2)+551) x+\frac{1}{36}
   (1505-288 \log(2))+\frac{2 (27 \log(2)-332)}{27 (1-x)}+\frac{2 (27
   \log(2)-332)}{27 x}\right)
\ebrk
+\ly^2 \left(\frac{x^2}{2}+\frac{503
   x}{36}+\pi ^2 \left(\frac{44 x^2}{3}-\frac{92 x}{3}-\frac{4}{3
   (1-x)}+\frac{101}{6}-\frac{23}{6 x}\right)-\frac{83}{8
   (1-x)}-\frac{347}{72}+\frac{1093}{36 x}-\frac{1}{x^2}\right)
\ebrk
+\s12
   \left(-4 x^2-\frac{61 x}{3}+\ly \left(-16 x^2+11
   x+\frac{11}{1-x}-65+\frac{30}{x}\right)+\lx \left(16 x^2-8
   x-\frac{34}{1-x}-38+\frac{82}{x}\right)
\brk
+\frac{23}{2
   (1-x)}+\frac{355}{6}-\frac{15}{x}\right)+\li3 \left(4 x^2-\frac{85
   x}{3}+\ly \left(-16 x^2+64
   x-\frac{4}{1-x}+10-\frac{44}{x}\right)
\brk
+\lx \left(52 x^2-58
   x+\frac{44}{1-x}-26-\frac{2}{x}\right)+\frac{15}{1-x}-\frac{209}{6}-\frac{23}{2
   x}\right)+\li2 \left(\left(-34 x^2+\frac{79
   x}{2}-\frac{7}{1-x}
\ibrk
-22+\frac{13}{2 x}\right) \lx^2+\left(-4
   x^2+\frac{85 x}{3}+\ly \left(16 x^2-64
   x+\frac{4}{1-x}-10+\frac{44}{x}\right)-\frac{15}{1-x}+\frac{209}{6}+\frac{23}{2
   x}\right) \lx
\brk
+\pi ^2 \left(42 x-\frac{34}{3
   (1-x)}-21+\frac{34}{3 x}\right)\right)+\left(-\frac{64 x^2}{3}+\frac{137
   x}{3}-\frac{69}{2 (1-x)}+\frac{51}{2}-\frac{8}{x}\right)
   \z3
\ebrk
+\ly \left(-8 x^2-\frac{461 x}{18}+\pi ^2 \left(\frac{16
   x^2}{3}-\frac{413 x}{18}-\frac{101}{18 (1-x)}+\frac{130}{9}+\frac{73}{18
   x}\right)+\left(24 x^2-43 x-\frac{113}{3 (1-x)}
\ibrk
+\frac{307}{3}-\frac{38}{3
   x}\right) \z3+\frac{839}{216 (1-x)}+\frac{3803}{216}+\frac{430}{27
   x}\right)-\frac{15317}{648 (1-x)}-\frac{15317}{648 x}+\frac{3353}{324},
\end{eqnarray}

\begin{eqnarray}
C &=& \frac{1}{\ep^3}\left\{
\lm\left[
\frac{2}{x}-4+\frac{2}{1-x}
\right]
+
\frac{2}{x}-4+\frac{2}{1-x}
\right\}
+
\frac{1}{\ep^2}\left\{
\lm^2\left[
-2 x^2+2 x+\frac{1}{2 (1-x)}-2+\frac{1}{2 x}
\right]
\brk
+
\lm \ls\left[
-\frac{4}{x}+8-\frac{4}{1-x}
\right]
+
\lm\left[
-4 x^2+4 x+\lx \left(-\frac{2}{x}+4-\frac{2}{1-x}\right)+\ly
   \left(-\frac{2}{x}+4-\frac{2}{1-x}\right)
\ibrk
+\frac{11}{2 (1-x)}-17+\frac{11}{2
   x}
\right]
+
\ls\left[
-\frac{4}{x}+8-\frac{4}{1-x}
\right]
+
\left(-\frac{1}{x}+1-\frac{2}{1-x}\right)
   \lx^2+\left(-\frac{5}{x}+5-\frac{2}{1-x}\right) \lx
\brk
-2
   x^2+\ly \left(-\frac{2}{x}+5-\frac{5}{1-x}\right)+\ly^2
   \left(-\frac{2}{x}+1-\frac{1}{1-x}\right)+2 x+\pi ^2 \left(2 x^2-2
   x-\frac{5}{6 (1-x)}+\frac{8}{3}-\frac{5}{6
   x}\right)
\brk
+\frac{6}{1-x}+\frac{6}{x}-17
\right\}
+
\frac{1}{\ep}\left\{
\lm^3\left[
2 x^2-2 x-\frac{7}{6 (1-x)}+\frac{10}{3}-\frac{7}{6 x}
\right]
+
\lm^2 \ls\left[
4 x^2-4 x-\frac{1}{1-x}+4-\frac{1}{x}
\right]
\brk
+
\lm \ls^2\left[
\frac{4}{x}-8+\frac{4}{1-x}
\right]
+
\lm^2\left[
2 x^2-2 x+\lx \left(\frac{1}{x}-2+\frac{1}{1-x}\right)+\ly
   \left(\frac{1}{x}-2+\frac{1}{1-x}\right)-\frac{10}{3
   (1-x)}
\ibrk
+\frac{23}{3}-\frac{10}{3 x}
\right]
+
\lm \ls\left[
8 x^2-8 x+\lx \left(\frac{4}{x}-8+\frac{4}{1-x}\right)+\ly
   \left(\frac{4}{x}-8+\frac{4}{1-x}\right)-\frac{11}{3
   (1-x)}+\frac{58}{3}-\frac{11}{3 x}
\right]
\brk
+
\ls^2\left[
\frac{4}{x}-8+\frac{4}{1-x}
\right]
+
\lm\left[
\left(-\frac{x}{2}-\frac{1}{2 x}\right) \lx^2+\left(\ly
   \left(\frac{2}{x}-4+\frac{2}{1-x}\right)-\frac{x}{2}+\frac{3}{1-x}-\frac{3}{2 x}+
   3\right) \lx
\ibrk
-6 x^2+\ly^2 \left(\frac{x}{2}-\frac{1}{2
   (1-x)}-\frac{1}{2}\right)+\ly \left(\frac{x}{2}-\frac{3}{2
   (1-x)}+\frac{5}{2}+\frac{3}{x}\right)+6 x
\ibrk
+\pi ^2 \left(\frac{8
   x^2}{3}-\frac{8 x}{3}-\frac{11}{6 (1-x)}+5-\frac{11}{6
   x}\right)-\frac{80}{9 (1-x)}-\frac{80}{9 x}+\frac{41}{18}
\right]
+
\ls\left[
\left(\frac{2}{x}-2+\frac{4}{1-x}\right)
   \lx^2
\ibrk
+\left(\frac{10}{x}-10+\frac{4}{1-x}\right) \lx+4
   x^2+\ly^2 \left(\frac{4}{x}-2+\frac{2}{1-x}\right)+\ly
   \left(\frac{4}{x}-10+\frac{10}{1-x}\right)-4 x
\ibrk
+\pi ^2 \left(-4 x^2+4
   x+\frac{5}{3 (1-x)}-\frac{16}{3}+\frac{5}{3 x}\right)-\frac{14}{3
   (1-x)}-\frac{14}{3 x}+\frac{58}{3}
\right]
+
\left(\frac{5}{3 x}-\frac{5}{3}+\frac{10}{3 (1-x)}\right)
   \lx^3
\brk
+\left(-\frac{x}{2}-\frac{5}{3 (1-x)}-\frac{5}{3}+\frac{25}{6
   x}\right) \lx^2+\li2 \left(-\frac{2}{x}+2-\frac{4}{1-x}\right)
   \lx+\left(\left(\frac{2}{x}-1+\frac{1}{1-x}\right)
   \ly^2
\ibrk
+\left(\frac{5}{x}-6+\frac{5}{1-x}\right)
   \ly-\frac{x}{2}+\pi ^2 \left(x-\frac{2}{3
   (1-x)}-\frac{5}{3}+\frac{4}{3
   x}\right)+\frac{1}{1-x}-\frac{10}{x}+\frac{77}{6}\right) \lx-6
   x^2
\brk
+\s12 \left(-\frac{4}{x}+2-\frac{2}{1-x}\right)+\li3
   \left(\frac{2}{x}-2+\frac{4}{1-x}\right)+\ly^3 \left(\frac{10}{3
   x}-\frac{5}{3}+\frac{5}{3 (1-x)}\right)
\brk
+\ly^2
   \left(\frac{x}{2}+\frac{25}{6 (1-x)}-\frac{13}{6}-\frac{5}{3
   x}\right)+\ly \left(\pi ^2 \left(-x+\frac{4}{3
   (1-x)}-\frac{2}{3}-\frac{2}{3
   x}\right)+\frac{x}{2}-\frac{10}{1-x}+\frac{1}{x}+\frac{37}{3}\right)
\brk
+6
   x+\pi ^2 \left(\frac{20 x^2}{3}-\frac{20 x}{3}-\frac{28}{9
   (1-x)}+\frac{139}{18}-\frac{28}{9
   x}\right)+\left(-\frac{7}{x}+12-\frac{9}{1-x}\right)
   \z3-\frac{121}{18 (1-x)}-\frac{121}{18 x}-\frac{73}{18}
\right\}
\ebrk
+
\lm^4\left[
-\frac{7 x^2}{6}+\frac{7 x}{6}+\frac{19}{24 (1-x)}-\frac{13}{6}+\frac{19}{24
   x}
\right]
+
\lm^3 \ls\left[
-4 x^2+4 x+\frac{7}{3 (1-x)}-\frac{20}{3}+\frac{7}{3 x}
\right]
\ebrk
+
\lm^2 \ls^2\left[
-4 x^2+4 x+\frac{1}{1-x}-4+\frac{1}{x}
\right]
+
\lm \ls^3\left[
-\frac{8}{3 x}+\frac{16}{3}-\frac{8}{3 (1-x)}
\right]
+
\lm^3\left[
-\frac{2 x^2}{3}+\frac{2 x}{3}
\brk
+\lx \left(-\frac{1}{3
   x}+\frac{2}{3}-\frac{1}{3 (1-x)}\right)+\ly \left(-\frac{1}{3
   x}+\frac{2}{3}-\frac{1}{3 (1-x)}\right)+\frac{3}{2 (1-x)}-2+\frac{3}{2 x}
\right]
\ebrk
+
\lm^2 \ls\left[
-4 x^2+4 x+\lx \left(-\frac{2}{x}+4-\frac{2}{1-x}\right)+\ly
   \left(-\frac{2}{x}+4-\frac{2}{1-x}\right)+\frac{3}{1-x}-8+\frac{3}{x}
\right]
\ebrk
+
\lm \ls^2\left[
-8 x^2+8 x+\lx \left(-\frac{4}{x}+8-\frac{4}{1-x}\right)+\ly
   \left(-\frac{4}{x}+8-\frac{4}{1-x}\right)-12
\right]
+
\ls^3\left[
-\frac{8}{3 x}+\frac{16}{3}
\brk
-\frac{8}{3 (1-x)}
\right]
+
\lm^2\left[
\left(\frac{x}{4}+\frac{1}{4 x}\right) \lx^2+\left(\ly
   \left(-\frac{1}{x}+2-\frac{1}{1-x}\right)+\frac{x}{4}-\frac{3}{2
   (1-x)}+\frac{3}{4 x}-\frac{3}{2}\right) \lx-\frac{7
   x^2}{2}
\brk
+\ly^2 \left(-\frac{x}{4}+\frac{1}{4
   (1-x)}+\frac{1}{4}\right)+\ly \left(-\frac{x}{4}+\frac{3}{4
   (1-x)}-\frac{5}{4}-\frac{3}{2 x}\right)+\frac{7 x}{2}+\pi ^2 \left(-\frac{5
   x^2}{3}+\frac{5 x}{3}+\frac{13}{12 (1-x)}
\ibrk
-3+\frac{13}{12
   x}\right)+\frac{739}{72 (1-x)}+\frac{739}{72 x}-\frac{643}{36}
\right]
+
\lm \ls\left[
\left(x+\frac{1}{x}\right) \lx^2+\left(\ly
   \left(-\frac{4}{x}+8-\frac{4}{1-x}\right)+x
\ibrk
-\frac{6}{1-x}+\frac{3}{x}-6\right)
   \lx+12 x^2+\ly^2 \left(-x+\frac{1}{1-x}+1\right)+\ly
   \left(-x+\frac{3}{1-x}-5-\frac{6}{x}\right)-12 x
\brk
+\pi ^2 \left(-\frac{16
   x^2}{3}+\frac{16 x}{3}+\frac{11}{3 (1-x)}-10+\frac{11}{3
   x}\right)+\frac{61}{9 (1-x)}+\frac{61}{9 x}+\frac{25}{9}
\right]
+
\ls^2\left[
\left(-\frac{2}{x}+2-\frac{4}{1-x}\right)
   \lx^2
\brk
+\left(-\frac{10}{x}+10-\frac{4}{1-x}\right) \lx-4
   x^2+\ly \left(-\frac{4}{x}+10-\frac{10}{1-x}\right)+\ly^2
   \left(-\frac{4}{x}+2-\frac{2}{1-x}\right)+4 x
\brk
+\pi ^2 \left(4 x^2-4
   x-\frac{5}{3 (1-x)}+\frac{16}{3}-\frac{5}{3
   x}\right)+\frac{1}{1-x}+\frac{1}{x}-12
\right]
+
\lm\left[
\left(\frac{x}{3}+\frac{1}{3 (1-x)}-\frac{2}{3}+\frac{2}{3 x}\right)
   \lx^3
\brk
+\left(\ly \left(-\frac{x}{2}-\frac{1}{2
   x}\right)+x+\frac{3}{x}-\frac{13}{2}\right)
   \lx^2+\left(\left(-\frac{1}{x}+2-\frac{1}{1-x}\right)
   \ly^2+\left(-\frac{3}{x}+8-\frac{3}{1-x}\right) \ly
\ibrk
+\pi ^2
   \left(\frac{x}{2}+\frac{1}{1-x}-2+\frac{3}{2 x}\right)-\frac{9
   x}{4}-\frac{4}{1-x}-\frac{31}{4 x}+15\right) \lx+\li2
   \left(-x+\frac{2}{1-x}-4+\frac{1}{x}\right) \lx-\frac{33 x^2}{2}
\brk
+\pi
   ^2 \left(\frac{8}{3 x}-\frac{25}{4}+\frac{8}{3 (1-x)}\right)+\ly^2
   \left(-x+\frac{3}{1-x}-\frac{11}{2}\right)+\ly^3
   \left(-\frac{x}{3}+\frac{2}{3 (1-x)}-\frac{1}{3}+\frac{1}{3
   x}\right)+\frac{33 x}{2}
\brk
+\li3
   \left(x-\frac{2}{1-x}+4-\frac{1}{x}\right)+\s12
   \left(x+\frac{1}{1-x}-5+\frac{2}{x}\right)+\ly \left(\pi ^2
   \left(-\frac{x}{2}+\frac{3}{2 (1-x)}-\frac{3}{2}+\frac{1}{x}\right)
\ibrk
+\frac{9
   x}{4}-\frac{31}{4 (1-x)}-\frac{4}{x}+\frac{51}{4}\right)+\left(32 x^2-33
   x-\frac{49}{3 (1-x)}+\frac{146}{3}-\frac{52}{3 x}\right)
   \z3+\frac{2797}{216 (1-x)}+\frac{2797}{216 x}
\brk
-\frac{2983}{108}
\right]
+
\ls\left[
\left(-\frac{10}{3 x}+\frac{10}{3}-\frac{20}{3 (1-x)}\right)
   \lx^3+\left(x-\frac{4}{1-x}+7-\frac{12}{x}\right)
   \lx^2+\li2 \left(\frac{4}{x}-4+\frac{8}{1-x}\right)
   \lx
\brk
+\left(\left(-\frac{4}{x}+2-\frac{2}{1-x}\right)
   \ly^2+\left(-\frac{10}{x}+12-\frac{10}{1-x}\right) \ly+\pi ^2
   \left(-2 x+\frac{4}{3 (1-x)}+\frac{10}{3}-\frac{8}{3
   x}\right)+x
\ibrk
-\frac{2}{1-x}+\frac{9}{x}-22\right) \lx+12
   x^2+\ly^3 \left(-\frac{20}{3 x}+\frac{10}{3}-\frac{10}{3
   (1-x)}\right)+\li3
   \left(-\frac{4}{x}+4-\frac{8}{1-x}\right)
\brk
+\s12
   \left(\frac{8}{x}-4+\frac{4}{1-x}\right)+\ly^2
   \left(-x-\frac{12}{1-x}+8-\frac{4}{x}\right)-12 x+\pi ^2 \left(-\frac{40
   x^2}{3}+\frac{40 x}{3}+\frac{4}{3 (1-x)}
\ibrk
-\frac{17}{3}+\frac{4}{3
   x}\right)+\ly \left(-x+\pi ^2 \left(2 x-\frac{8}{3
   (1-x)}+\frac{4}{3}+\frac{4}{3
   x}\right)+\frac{9}{1-x}-21-\frac{2}{x}\right)+\left(\frac{14}{x}-24+
   \frac{18}{1-x}\right) \z3
\brk
+\frac{88}{9 (1-x)}+\frac{88}{9 x}+\frac{7}{9}
\right]
+
\left(\frac{5 x}{24}-\frac{17}{6 (1-x)}+\frac{3}{2}-\frac{25}{24 x}\right)
   \lx^4+\left(\ly \left(\frac{1}{3 x}-\frac{1}{3}+\frac{2}{3
   (1-x)}\right)-\frac{3 x}{4}
\brk
-\frac{5}{6 (1-x)}-\frac{47}{12
   x}+\frac{17}{6}\right) \lx^3+\left(\left(-\frac{3
   x}{2}+\frac{5}{2}-\frac{9}{2 x}\right) \ly^2+\left(\frac{23
   x}{4}-\frac{31}{6 (1-x)}-\frac{8}{3}-\frac{37}{12 x}\right)
   \ly+\frac{11 x^2}{2}
\brk
+\pi ^2 \left(-\frac{x}{3}-\frac{5}{6
   (1-x)}+1-\frac{1}{6 x}\right)-\frac{49 x}{8}+\frac{91}{9
   (1-x)}+\frac{991}{72 x}+\frac{1}{(x-1)^2}-\frac{128}{9}\right)
   \lx^2
\ebrk
+\left(\left(\frac{x}{3}-\frac{5}{3
   (1-x)}-\frac{1}{3}-\frac{5}{3 x}\right) \ly^3+\left(-\frac{7
   x}{2}-\frac{15}{2 (1-x)}+8-\frac{6}{x}\right) \ly^2+\left(-11 x^2+11
   x
\ibrk
+\pi ^2 \left(\frac{4 x}{3}-\frac{26}{3
   (1-x)}+\frac{23}{3}-\frac{4}{x}\right)-12\right) \ly+\pi ^2 \left(-2
   x+\frac{11}{9 (1-x)}-\frac{53}{18}+\frac{35}{18 x}\right)-\frac{29
   x}{8}
\brk
+\left(\frac{10}{x}-28+\frac{16}{1-x}\right)
   \z3+\frac{8}{1-x}+\frac{103}{24 x}+\frac{473}{18}\right) \lx-38
   x^2+\s12 \left(\lx \left(-6
   x-\frac{12}{1-x}+12-\frac{12}{x}\right)
\brk
+\ly
   \left(-x+\frac{7}{1-x}+7+\frac{10}{x}\right)-\frac{9 x}{2}+\frac{53}{6
   (1-x)}+\frac{5}{3 x}-\frac{59}{6}\right)+\li4
   \left(-x-\frac{16}{1-x}-20-\frac{9}{x}\right)
\ebrk
+\s13
   \left(-x+\frac{9}{1-x}+21+\frac{16}{x}\right)+\ly^4 \left(-\frac{5
   x}{24}-\frac{25}{24 (1-x)}+\frac{41}{24}-\frac{17}{6 x}\right)+\ly^3
   \left(\frac{3 x}{4}-\frac{47}{12 (1-x)}
\brk
+\frac{25}{12}-\frac{5}{6
   x}\right)+38 x+\s22 \left(12 x-\frac{6}{1-x}-6+\frac{6}{x}\right)+\pi
   ^4 \left(-\frac{53 x^2}{45}+\frac{19 x}{20}+\frac{217}{360
   (1-x)}+\frac{199}{180}
\brk
-\frac{97}{360 x}\right)+\ly^2 \left(\frac{11
   x^2}{2}-\frac{39 x}{8}+\pi ^2 \left(\frac{x}{3}-\frac{1}{6
   (1-x)}+\frac{2}{3}-\frac{5}{6 x}\right)+\frac{991}{72
   (1-x)}-\frac{1069}{72}+\frac{91}{9 x}+\frac{1}{x^2}\right)
\ebrk
+\pi ^2
   \left(\frac{1}{6} (48 \log(2)+107) x^2-\frac{1}{6} (48 \log(2)+107)
   x+\frac{1}{54} (648 \log(2)-1175)-\frac{432 \log(2)-1379}{108
   (1-x)}
\brk
+\frac{1379-432 \log(2)}{108 x}\right)+\li2
   \left(\left(\frac{x}{2}+\frac{2}{1-x}-4+\frac{5}{2 x}\right)
   \lx^2+\left(\ly \left(-6
   x-\frac{6}{1-x}+12-\frac{18}{x}\right)
\ibrk
+\frac{9 x}{2}+\frac{5}{3
   (1-x)}+\frac{53}{6 x}-\frac{43}{3}\right) \lx+\pi ^2 \left(\frac{8
   x}{3}-\frac{14}{3 (1-x)}-\frac{4}{3}+\frac{14}{3
   x}\right)\right)+\li3 \left(\lx
   \left(\frac{2}{x}+14+\frac{6}{1-x}\right)
\brk
-\frac{9 x}{2}+\ly \left(6
   x+\frac{6}{1-x}-12+\frac{18}{x}\right)-\frac{5}{3 (1-x)}-\frac{53}{6
   x}+\frac{43}{3}\right)+\left(44 x^2-\frac{79 x}{2}-\frac{139}{6
   (1-x)}+\frac{383}{6}-\frac{16}{x}\right) \z3
\ebrk
+\ly \left(\frac{29
   x}{8}+\pi ^2 \left(2 x+\frac{35}{18 (1-x)}-\frac{89}{18}+\frac{11}{9
   x}\right)+\left(-5 x-\frac{3}{1-x}-23-\frac{12}{x}\right)
   \z3+\frac{103}{24
   (1-x)}+\frac{1631}{72}
\brk
+\frac{8}{x}\right)+\frac{937}{216
   (1-x)}+\frac{937}{216 x}-\frac{5369}{216},
\end{eqnarray}

\begin{eqnarray}
D &=& \frac{1}{\ep^2}\left\{
\lm^2\left[
-\frac{1}{2 x}+1-\frac{1}{2 (1-x)}
\right]
+
\lm\left[
-\frac{1}{x}+2-\frac{1}{1-x}
\right]
+
\pi ^2 \left(\frac{1}{2 x}-1+\frac{1}{2 (1-x)}\right)-\frac{1}{2
   (1-x)}
\brk
-\frac{1}{2 x}+1
\right\}
+
\frac{1}{\ep}\left\{
\lm^3\left[
\frac{1}{2 x}-1+\frac{1}{2 (1-x)}
\right]
+
\lm^2 \ls\left[
\frac{1}{x}-2+\frac{1}{1-x}
\right]
+
\lm^2\left[
\frac{1}{x}-1+\frac{1}{1-x}
\right]
\brk
+
\lm \ls\left[
\frac{2}{x}-4+\frac{2}{1-x}
\right]
+
\lm\left[
\left(\frac{1}{2 x}-\frac{1}{2}+\frac{1}{1-x}\right)
   \lx^2+\left(\frac{3}{2 x}-\frac{1}{2}\right) \lx+\ly
   \left(\frac{3}{2 (1-x)}-\frac{1}{2}\right)
\ibrk
+\pi ^2 \left(\frac{2}{3
   x}-\frac{4}{3}+\frac{2}{3 (1-x)}\right)+\ly^2
   \left(\frac{1}{x}-\frac{1}{2}+\frac{1}{2 (1-x)}\right)+2
\right]
+
\ls\left[
\pi ^2 \left(-\frac{1}{x}+2-\frac{1}{1-x}\right)+\frac{1}{1-x}
\ibrk
+\frac{1}{x}-2
\right]
+
\left(\frac{1}{2 x}-\frac{1}{2}+\frac{1}{1-x}\right) \lx^2+\left(\pi ^2
   \left(-\frac{1}{x}+1-\frac{2}{1-x}\right)+\frac{3}{2 x}-\frac{1}{2}\right)
   \lx+\ly \left(\pi ^2
   \left(-\frac{2}{x}+1
\iibrk
-\frac{1}{1-x}\right)+\frac{3}{2
   (1-x)}-\frac{1}{2}\right)+\pi ^2 \left(-\frac{1}{3
   x}-\frac{7}{3}-\frac{1}{3 (1-x)}\right)+\ly^2
   \left(\frac{1}{x}-\frac{1}{2}+\frac{1}{2 (1-x)}\right)-\frac{1}{2
   (1-x)}
\brk
-\frac{1}{2 x}+2
\right\}
+
\lm^4\left[
-\frac{7}{24 x}+\frac{7}{12}-\frac{7}{24 (1-x)}
\right]
+
\lm^3 \ls\left[
-\frac{1}{x}+2-\frac{1}{1-x}
\right]
+
\lm^2 \ls^2\left[
-\frac{1}{x}+2-\frac{1}{1-x}
\right]
\ebrk
+
\lm^3\left[
-\frac{2}{3 x}+\frac{1}{3}-\frac{2}{3 (1-x)}
\right]
+
\lm^2 \ls\left[
-\frac{2}{x}+2-\frac{2}{1-x}
\right]
+
\lm \ls^2\left[
-\frac{2}{x}+4-\frac{2}{1-x}
\right]
\ebrk
+
\lm^2\left[
\left(-\frac{1}{4 x}+\frac{1}{4}-\frac{1}{2 (1-x)}\right)
   \lx^2+\left(\frac{1}{4}-\frac{3}{4 x}\right) \lx+\ly
   \left(\frac{1}{4}-\frac{3}{4 (1-x)}\right)+\ly^2 \left(-\frac{1}{2
   x}+\frac{1}{4}-\frac{1}{4 (1-x)}\right)
\brk
+\pi ^2 \left(-\frac{5}{12
   x}+\frac{5}{6}-\frac{5}{12 (1-x)}\right)-\frac{13}{8 (1-x)}-\frac{13}{8
   x}+\frac{9}{4}
\right]
+
\lm \ls\left[
\left(-\frac{1}{x}+1-\frac{2}{1-x}\right)
   \lx^2+\left(1-\frac{3}{x}\right) \lx
\brk
+\ly
   \left(1-\frac{3}{1-x}\right)+\ly^2
   \left(-\frac{2}{x}+1-\frac{1}{1-x}\right)+\pi ^2 \left(-\frac{4}{3
   x}+\frac{8}{3}-\frac{4}{3 (1-x)}\right)-4
\right]
+
\ls^2\left[
\pi ^2 \left(\frac{1}{x}-2
\ibrk
+\frac{1}{1-x}\right)-\frac{1}{1-x}-\frac{1}{x}+2
\right]
+
\lm\left[
\left(-\frac{1}{3 x}+\frac{1}{3}-\frac{2}{3 (1-x)}\right)
   \lx^3+\left(\ly \left(\frac{1}{2
   x}-\frac{1}{2}+\frac{1}{1-x}\right)-\frac{1}{2 (1-x)}
\ibrk
-\frac{3}{2
   x}+1\right) \lx^2+\li2 \left(\frac{1}{x}-1+\frac{2}{1-x}\right)
   \lx+\left(\pi ^2 \left(-\frac{1}{2
   x}+\frac{1}{2}-\frac{1}{1-x}\right)+\frac{5}{4 x}-\frac{9}{4}\right)
   \lx
\brk
+\ly \left(\pi ^2 \left(-\frac{1}{x}+\frac{1}{2}-\frac{1}{2
   (1-x)}\right)+\frac{5}{4 (1-x)}-\frac{9}{4}\right)+\pi ^2
   \left(-\frac{17}{12 x}+\frac{1}{2}-\frac{17}{12 (1-x)}\right)
\brk
+\li3
   \left(-\frac{1}{x}+1-\frac{2}{1-x}\right)+\ly^3 \left(-\frac{2}{3
   x}+\frac{1}{3}-\frac{1}{3 (1-x)}\right)+\ly^2 \left(-\frac{1}{2
   x}+1-\frac{3}{2 (1-x)}\right)
\brk
+\s12
   \left(\frac{2}{x}-1+\frac{1}{1-x}\right)+\left(\frac{9}{x}-17+\frac{10}{1-x
   }\right) \z3-\frac{11}{8 (1-x)}-\frac{11}{8 x}+\frac{19}{4}
\right]
+
\ls\left[
\left(-\frac{1}{x}+1
\ibrk
-\frac{2}{1-x}\right) \lx^2+\left(\pi ^2
   \left(\frac{2}{x}-2+\frac{4}{1-x}\right)-\frac{3}{x}+1\right)
   \lx+\ly \left(\pi ^2
   \left(\frac{4}{x}-2+\frac{2}{1-x}\right)-\frac{3}{1-x}+1\right)
\brk
+\ly^2
   \left(-\frac{2}{x}+1-\frac{1}{1-x}\right)+\pi ^2 \left(\frac{2}{3
   x}+\frac{14}{3}+\frac{2}{3 (1-x)}\right)+\frac{1}{1-x}+\frac{1}{x}-4
\right]
+
\left(-\frac{1}{8 x}+\frac{5}{24}-\frac{2}{3 (1-x)}\right)
   \lx^4
\ebrk
+\left(-\frac{13}{12 x}-\frac{3}{4}-\frac{2}{3 (1-x)}\right)
   \lx^3+\left(\left(-\frac{4}{x}+\frac{1}{2}-\frac{1}{2 (1-x)}\right)
   \ly^2+\left(\frac{5}{4 x}-\frac{13}{4}-\frac{9}{2 (1-x)}\right)
   \ly+3 x^2
\brk
+\pi ^2 \left(\frac{2}{3 x}-\frac{1}{3}-\frac{1}{3
   (1-x)}\right)-3 x-\frac{5}{4 (1-x)}-\frac{17}{8
   x}-\frac{1}{(x-1)^2}+\frac{43}{8}\right) \lx^2+\left(\left(\frac{5}{3
   x}-\frac{1}{3}\right) \ly^3
\brk
+\left(\frac{7}{2}-\frac{3}{2 x}\right)
   \ly^2+\left(-6 x^2+6 x+\pi ^2 \left(-\frac{5}{3
   x}+\frac{4}{3}-\frac{5}{3 (1-x)}\right)-10\right) \ly+\pi ^2
   \left(\frac{1}{x}-\frac{11}{3}+\frac{4}{3 (1-x)}\right)
\brk
-6
   x+\left(\frac{6}{1-x}-\frac{2}{x}\right)
   \z3+\frac{1}{1-x}+\frac{19}{8 x}+\frac{27}{8}\right)
   \lx+\li2 \left(\left(-\frac{1}{2
   x}+\frac{1}{2}-\frac{1}{1-x}\right) \lx^2
\brk
+\left(\ly
   \left(-\frac{16}{x}+2-\frac{2}{1-x}\right)-\frac{6}{1-x}+\frac{5}{2
   x}-\frac{27}{2}\right) \lx\right)+\s13
   \left(-\frac{14}{x}+1+\frac{3}{1-x}\right)
\ebrk
+\s12 \left(\lx
   \left(-\frac{16}{x}+2-\frac{2}{1-x}\right)+\ly
   \left(-\frac{8}{x}+1+\frac{1}{1-x}\right)+\frac{5}{2
   (1-x)}-\frac{6}{x}-\frac{27}{2}\right)
\ebrk
+\li4
   \left(-\frac{3}{x}-1+\frac{14}{1-x}\right)+\li3 \left(\lx
   \left(\frac{2}{x}-\frac{6}{1-x}\right)+\ly
   \left(\frac{16}{x}-2+\frac{2}{1-x}\right)+\frac{6}{1-x}-\frac{5}{2
   x}+\frac{27}{2}\right)
\ebrk
+\ly^3 \left(-\frac{2}{3
   x}-\frac{3}{4}-\frac{13}{12 (1-x)}\right)+\ly^4 \left(-\frac{2}{3
   x}+\frac{5}{24}-\frac{1}{8 (1-x)}\right)+\pi ^4 \left(-\frac{1}{30
   x}+\frac{29}{60}-\frac{11}{60 (1-x)}\right)
\ebrk
+\s22
   \left(\frac{14}{x}-\frac{14}{1-x}\right)+\ly^2 \left(3 x^2-3 x+\pi ^2
   \left(-\frac{1}{3 x}-\frac{1}{3}+\frac{2}{3 (1-x)}\right)-\frac{17}{8
   (1-x)}+\frac{43}{8}-\frac{5}{4 x}-\frac{1}{x^2}\right)
\ebrk
+\pi ^2 \left(3 x^2-3
   x+\frac{1}{2} (13-8 \log(2))-\frac{13-24 \log(2)}{12 (1-x)}+\frac{24
   \log(2)-13}{12 x}\right)+\left(\frac{15}{x}-\frac{9}{2}+\frac{13}{2
   (1-x)}\right) \z3
\ebrk
+\ly \left(\pi ^2 \left(\frac{4}{3
   x}-\frac{11}{3}+\frac{1}{1-x}\right)+6
   x+\left(-\frac{2}{x}+1-\frac{5}{1-x}\right) \z3+\frac{19}{8
   (1-x)}+\frac{1}{x}-\frac{21}{8}\right)-\frac{47}{8 (1-x)}
\ebrk
-\frac{47}{8
   x}+\frac{57}{4},
\end{eqnarray}

\begin{eqnarray}
E_l &=& \frac{1}{\ep^3}\left\{
\frac{28 x^2}{3}-\frac{28 x}{3}-\frac{7}{3 (1-x)}+\frac{28}{3}-\frac{7}{3 x}
\right\}
+
\frac{1}{\ep^2}\left\{
\lm\left[
4 x^2-4 x-\frac{1}{1-x}+4-\frac{1}{x}
\right]
+
\ls\left[
-8 x^2+8 x
\ibrk
+\frac{2}{1-x}-8+\frac{2}{x}
\right]
+
12 x^2-12 x+\ly \left(-4 x^2+\frac{2}{1-x}-2\right)+\lx \left(-4
   x^2+8 x-6+\frac{2}{x}\right)-\frac{2}{3 (1-x)}
\brk
+\frac{22}{3}-\frac{2}{3 x}
\right\}
+
\frac{1}{\ep}\left\{
\lm^2\left[
-\frac{4 x^2}{3}+\frac{4 x}{3}+\frac{1}{3 (1-x)}-\frac{4}{3}+\frac{1}{3 x}
\right]
+
\lm \ls\left[
-\frac{8 x^2}{3}+\frac{8 x}{3}+\frac{2}{3 (1-x)}-\frac{8}{3}+\frac{2}{3 x}
\right]
\brk
+
\ls^2\left[
\frac{8 x^2}{3}-\frac{8 x}{3}-\frac{2}{3 (1-x)}+\frac{8}{3}-\frac{2}{3 x}
\right]
+
\lm\left[
-\frac{44 x^2}{9}+\frac{44 x}{9}+\frac{20}{9 (1-x)}-\frac{62}{9}+\frac{20}{9
   x}
\right]
+
\ls\left[
\frac{88 x^2}{9}
\ibrk
-\frac{88 x}{9}+\ly \left(\frac{8 x^2}{3}-\frac{4}{3
   (1-x)}+\frac{4}{3}\right)+\lx \left(\frac{8 x^2}{3}-\frac{16
   x}{3}+4-\frac{4}{3 x}\right)-\frac{40}{9 (1-x)}+\frac{124}{9}-\frac{40}{9
   x}
\right]
\brk
+
\left(-\frac{4 x}{3}+1-\frac{1}{3 x}\right) \lx^2+\left(\frac{56
   x^2}{9}-\frac{97 x}{9}+\frac{26}{3}-\frac{37}{9 x}\right)
   \lx-\frac{802 x^2}{27}+\frac{802 x}{27}+\ly^2 \left(\frac{4
   x}{3}-\frac{1}{3 (1-x)}-\frac{1}{3}\right)
\brk
+\pi ^2 \left(-2 x^2+2
   x+\frac{1}{2 (1-x)}-2+\frac{1}{2 x}\right)+\ly \left(\frac{56
   x^2}{9}-\frac{5 x}{3}-\frac{37}{9 (1-x)}+\frac{37}{9}\right)+\frac{581}{54
   (1-x)}+\frac{581}{54 x}
\brk
-\frac{991}{27}
\right\}
+
\lm^2\left[
\frac{32 x^2}{9}-\frac{32 x}{9}-\frac{11}{9 (1-x)}+\frac{38}{9}-\frac{11}{9 x}
\right]
+
\lm \ls\left[
\frac{64 x^2}{9}-\frac{64 x}{9}-\frac{22}{9 (1-x)}+\frac{76}{9}-\frac{22}{9 x}
\right]
\ebrk
+
\ls^2\left[
-\frac{64 x^2}{9}+\frac{64 x}{9}+\frac{22}{9 (1-x)}-\frac{76}{9}+\frac{22}{9
   x}
\right]
+
\lm\left[
\frac{50 x^2}{27}-\frac{50 x}{27}+\pi ^2 \left(-\frac{2 x^2}{3}+\frac{2
   x}{3}+\frac{1}{6 (1-x)}-\frac{2}{3}
\ibrk
+\frac{1}{6 x}\right)-\frac{73}{54
   (1-x)}+\frac{89}{27}-\frac{73}{54 x}
\right]
+
\ls\left[
\frac{80 x^2}{27}-\frac{80 x}{27}+\ly \left(-\frac{64
   x^2}{9}+\frac{44}{9 (1-x)}-\frac{44}{9}\right)+\lx \left(-\frac{64
   x^2}{9}
\ibrk
+\frac{128 x}{9}-12+\frac{44}{9 x}\right)+\pi ^2 \left(-\frac{4
   x^2}{9}+\frac{4 x}{9}+\frac{1}{9 (1-x)}-\frac{4}{9}+\frac{1}{9
   x}\right)+\frac{10}{27 (1-x)}+\frac{56}{27}+\frac{10}{27 x}
\right]
\ebrk
+
\left(\left(-x+3-\frac{2}{x}\right) \ly^2+\left(\frac{8
   x}{3}-3-\frac{1}{3 x}\right)
   \ly-\frac{x}{9}+\frac{2}{1-x}+\frac{11}{9 x}-\frac{10}{3}\right)
   \lx^2+\left(\frac{16 x^2}{27}-\frac{133 x}{54}
\brk
+\ly \pi ^2
   \left(\frac{2 x}{3}-2+\frac{4}{3 x}\right)+\pi ^2 \left(\frac{2
   x^2}{9}-\frac{4 x}{3}+\frac{4}{3}\right)+\frac{47}{9}+\frac{35}{54
   x}\right) \lx+\frac{2170 x^2}{81}
\ebrk
+\s12 \left(\lx \left(-4
   x+12-\frac{8}{x}\right)-\frac{16 x}{3}-\frac{2}{3
   (1-x)}-\frac{2}{3}\right)+\pi ^4 \left(-\frac{11 x}{90}+\frac{11}{45
   (1-x)}-\frac{11}{45}\right)
\ebrk
+\ly^2 \left(\frac{x}{9}+\frac{11}{9
   (1-x)}-\frac{31}{9}+\frac{2}{x}\right)-\frac{2170 x}{81}+\s22 \left(8
   x-\frac{8}{1-x}-4+\frac{8}{x}\right)+\pi ^2 \left(\frac{146
   x^2}{27}-\frac{146 x}{27}
\brk
-\frac{109}{54 (1-x)}+\frac{307}{54}-\frac{109}{54
   x}\right)+\li3 \left(-\frac{16 x}{3}+\ly \left(4
   x-12+\frac{8}{x}\right)+6+\frac{2}{3 x}\right)
\ebrk
+\li2 \left(\pi ^2
   \left(\frac{4 x}{3}-\frac{4}{3 (1-x)}-\frac{2}{3}+\frac{4}{3
   x}\right)+\lx \left(\ly \left(-4
   x+12-\frac{8}{x}\right)+\frac{16 x}{3}-\frac{2}{3
   x}-6\right)\right)
\ebrk
+\left(-8 x^2+\frac{40 x}{3}+\frac{2}{1-x}-14+\frac{4}{3
   x}\right) \z3+\ly \left(\frac{16 x^2}{27}+\frac{23 x}{18}+\pi
   ^2 \left(\frac{2 x^2}{9}+\frac{8 x}{9}+\frac{2}{9}\right)+\left(-4
   x+12-\frac{8}{x}\right) \z3
\brk
+\frac{35}{54
   (1-x)}+\frac{181}{54}\right)-\frac{1487}{162 (1-x)}-\frac{1487}{162
   x}+\frac{4289}{162},
\end{eqnarray}

\begin{eqnarray}
E_h &=& \frac{1}{\ep^2}\left\{
\lm\left[
\frac{32 x^2}{3}-\frac{32 x}{3}-\frac{8}{3 (1-x)}+\frac{32}{3}-\frac{8}{3 x}
\right]
+
\ls\left[
\frac{32 x^2}{3}-\frac{32 x}{3}-\frac{8}{3 (1-x)}+\frac{32}{3}-\frac{8}{3 x}
\right]
\right\}
\ebrk
+
\frac{1}{\ep}\left\{
\lm \ls\left[
-16 x^2+16 x+\frac{4}{1-x}-16+\frac{4}{x}
\right]
+
\ls^2\left[
-16 x^2+16 x+\frac{4}{1-x}-16+\frac{4}{x}
\right]
\brk
+
\lm\left[
\frac{128 x^2}{9}-\frac{128 x}{9}+\ly \left(-\frac{16 x^2}{3}+\frac{8}{3
   (1-x)}-\frac{8}{3}\right)+\lx \left(-\frac{16 x^2}{3}+\frac{32
   x}{3}-8+\frac{8}{3 x}\right)-\frac{8}{9 (1-x)}
\ibrk
+\frac{80}{9}-\frac{8}{9 x}
\right]
+
\ls\left[
\frac{128 x^2}{9}-\frac{128 x}{9}+\ly \left(-\frac{16 x^2}{3}+\frac{8}{3
   (1-x)}-\frac{8}{3}\right)+\lx \left(-\frac{16 x^2}{3}+\frac{32
   x}{3}-8+\frac{8}{3 x}\right)
\ibrk
-\frac{8}{9 (1-x)}+\frac{80}{9}-\frac{8}{9 x}
\right]
+
\frac{8 x^2}{3}-\frac{8 x}{3}+\pi ^2 \left(-\frac{8 x^2}{9}+\frac{8
   x}{9}+\frac{2}{9 (1-x)}-\frac{8}{9}+\frac{2}{9 x}\right)+\frac{2}{3}
\right\}
\ebrk
+
\lm^3\left[
-\frac{44 x^2}{9}+\frac{44 x}{9}+\frac{11}{9 (1-x)}-\frac{44}{9}+\frac{11}{9
   x}
\right]
+
\lm^2 \ls\left[
-\frac{8 x^2}{3}+\frac{8 x}{3}+\frac{2}{3 (1-x)}-\frac{8}{3}+\frac{2}{3 x}
\right]
\ebrk
+
\lm \ls^2\left[
\frac{40 x^2}{3}-\frac{40 x}{3}-\frac{10}{3 (1-x)}+\frac{40}{3}-\frac{10}{3 x}
\right]
+
\ls^3\left[
\frac{112 x^2}{9}-\frac{112 x}{9}-\frac{28}{9 (1-x)}+\frac{112}{9}-\frac{28}{9
   x}
\right]
\ebrk
+
\lm^2\left[
-\frac{166 x^2}{9}+\frac{166 x}{9}+\ly \left(4
   x^2-\frac{2}{1-x}+2\right)+\lx \left(4 x^2-8
   x+6-\frac{2}{x}\right)+\frac{83}{18 (1-x)}-\frac{166}{9}+\frac{83}{18 x}
\right]
\ebrk
+
\lm \ls\left[
-\frac{104 x^2}{9}+\frac{104 x}{9}+\ly \left(\frac{32
   x^2}{3}-\frac{16}{3 (1-x)}+\frac{16}{3}\right)+\lx \left(\frac{32
   x^2}{3}-\frac{64 x}{3}+16-\frac{16}{3 x}\right)-\frac{10}{9
   (1-x)}
\brk
-\frac{32}{9}-\frac{10}{9 x}
\right]
+
\ls^2\left[
-\frac{16 x^2}{9}+\frac{16 x}{9}+\ly \left(8
   x^2-\frac{4}{1-x}+4\right)+\lx \left(8 x^2-16
   x+12-\frac{4}{x}\right)-\frac{32}{9 (1-x)}
\brk
+\frac{56}{9}-\frac{32}{9 x}
\right]
+
\lm\left[
\left(-\frac{4 x}{3}+1-\frac{1}{3 x}\right) \lx^2+\left(\frac{88
   x^2}{9}-\frac{161 x}{9}+\frac{44}{3}-\frac{59}{9 x}\right)
   \lx-\frac{598 x^2}{9}+\frac{598 x}{9}
\brk
+\ly^2 \left(\frac{4
   x}{3}-\frac{1}{3 (1-x)}-\frac{1}{3}\right)+\pi ^2 \left(-\frac{20
   x^2}{9}+\frac{20 x}{9}+\frac{5}{9 (1-x)}-\frac{20}{9}+\frac{5}{9
   x}\right)+\ly \left(\frac{88 x^2}{9}-\frac{5 x}{3}
\ibrk
-\frac{59}{9
   (1-x)}+\frac{59}{9}\right)+\frac{121}{6 (1-x)}+\frac{121}{6
   x}-\frac{662}{9}
\right]
+
\ls\left[
\left(-\frac{8 x}{3}+2-\frac{2}{3 x}\right) \lx^2+\left(\frac{16
   x^2}{3}-\frac{22 x}{3}+\frac{16}{3}-\frac{10}{3 x}\right)
   \lx
\brk
-\frac{380 x^2}{9}+\frac{380 x}{9}+\ly^2 \left(\frac{8
   x}{3}-\frac{2}{3 (1-x)}-\frac{2}{3}\right)+\pi ^2 \left(-\frac{8
   x^2}{3}+\frac{8 x}{3}+\frac{2}{3 (1-x)}-\frac{8}{3}+\frac{2}{3
   x}\right)
\brk
+\ly \left(\frac{16 x^2}{3}-\frac{10 x}{3}-\frac{10}{3
   (1-x)}+\frac{10}{3}\right)+\frac{109}{9 (1-x)}+\frac{109}{9
   x}-\frac{130}{3}
\right]
-\frac{4}{9} x^2 \lx^3+\left(\left(-x+3-\frac{2}{x}\right)
   \ly^2
\brk
+\left(\frac{4 x^2}{3}+\frac{4 x}{3}-2-\frac{2}{3 x}\right)
   \ly-\frac{11 x}{18}+\frac{2}{1-x}+\frac{19}{18 x}-\frac{13}{3}\right)
   \lx^2+\left(\frac{188 x^2}{27}-\frac{893 x}{54}+\ly \pi ^2
   \left(\frac{2 x}{3}-2+\frac{4}{3 x}\right)
\brk
+\pi ^2 \left(-\frac{8
   x^2}{9}-\frac{4 x}{9}+\frac{2}{3}+\frac{2}{9
   x}\right)+\frac{176}{9}-\frac{323}{54 x}\right) \lx-\frac{3485
   x^2}{81}+\pi ^4 \left(-\frac{11 x}{90}+\frac{11}{45
   (1-x)}-\frac{11}{45}\right)
\ebrk
+\ly^2 \left(\frac{11 x}{18}+\frac{19}{18
   (1-x)}-\frac{89}{18}+\frac{2}{x}\right)+\frac{3485 x}{81}+\s22
   \left(8 x-\frac{8}{1-x}-4+\frac{8}{x}\right)+\ly^3 \left(-\frac{4
   x^2}{9}+\frac{8 x}{9}-\frac{4}{9}\right)
\ebrk
+\s12 \left(\frac{8 x^2}{3}-8
   x+\lx \left(-4 x+12-\frac{8}{x}\right)-\frac{4}{3
   (1-x)}+\frac{4}{3}\right)+\pi ^2 \left(\frac{140 x^2}{27}-\frac{140
   x}{27}-\frac{32}{27 (1-x)}
\brk
+\frac{83}{27}-\frac{32}{27 x}\right)+\li3
   \left(-\frac{8 x^2}{3}-\frac{8 x}{3}+\ly \left(4
   x-12+\frac{8}{x}\right)+4+\frac{4}{3 x}\right)+\li2 \left(\pi ^2
   \left(\frac{4 x}{3}
\ibrk
-\frac{4}{3 (1-x)}-\frac{2}{3}+\frac{4}{3
   x}\right)+\lx \left(\frac{8 x^2}{3}+\frac{8 x}{3}+\ly \left(-4
   x+12-\frac{8}{x}\right)-4-\frac{4}{3 x}\right)\right)+\left(\frac{152
   x^2}{9}-\frac{104 x}{9}
\brk
-\frac{32}{9 (1-x)}+\frac{92}{9}-\frac{44}{9
   x}\right) \z3+\ly \left(\frac{188 x^2}{27}+\frac{47 x}{18}+\pi
   ^2 \left(-\frac{8 x^2}{9}+\frac{20 x}{9}+\frac{2}{9
   (1-x)}-\frac{2}{3}\right)
\brk
+\left(-4 x+12-\frac{8}{x}\right)
   \z3-\frac{323}{54 (1-x)}+\frac{539}{54}\right)+\frac{5447}{324
   (1-x)}+\frac{5447}{324 x}-\frac{9913}{162},
\end{eqnarray}

\begin{eqnarray}
F_l &=& \frac{1}{\ep^3}\left\{
\frac{7}{3 x}-\frac{14}{3}+\frac{7}{3 (1-x)}
\right\}
+
\frac{1}{\ep^2}\left\{
\lm\left[
-4 x^2+4 x+\frac{2}{1-x}-6+\frac{2}{x}
\right]
+
\ls\left[
-\frac{2}{x}+4-\frac{2}{1-x}
\right]
-4 x^2
\brk
+4 x+\lx \left(-\frac{2}{x}+4-\frac{2}{1-x}\right)+\ly
   \left(-\frac{2}{x}+4-\frac{2}{1-x}\right)+\frac{5}{3 (1-x)}-10+\frac{5}{3
   x}
\right\}
\ebrk
+
\frac{1}{\ep}\left\{
\lm^2\left[
\frac{4 x^2}{3}-\frac{4 x}{3}-\frac{2}{3 (1-x)}+2-\frac{2}{3 x}
\right]
+
\lm \ls\left[
\frac{8 x^2}{3}-\frac{8 x}{3}-\frac{4}{3 (1-x)}+4-\frac{4}{3 x}
\right]
+
\ls^2\left[
\frac{2}{3 x}-\frac{4}{3}
\ibrk
+\frac{2}{3 (1-x)}
\right]
+
\lm\left[
\frac{44 x^2}{9}-\frac{44 x}{9}-\frac{40}{9 (1-x)}+\frac{28}{3}-\frac{40}{9 x}
\right]
+
\ls\left[
\frac{8 x^2}{3}-\frac{8 x}{3}+\lx \left(\frac{4}{3
   x}-\frac{8}{3}+\frac{4}{3 (1-x)}\right)
\ibrk
+\ly \left(\frac{4}{3
   x}-\frac{8}{3}+\frac{4}{3 (1-x)}\right)+\frac{34}{9
   (1-x)}-\frac{20}{9}+\frac{34}{9 x}
\right]
+
\left(-\frac{x}{3}+\frac{2}{3 (1-x)}-\frac{1}{3}\right)
   \lx^2
\brk
+\left(\ly \left(\frac{4}{3 x}-\frac{8}{3}+\frac{4}{3
   (1-x)}\right)-\frac{x}{3}+\frac{46}{9 (1-x)}+\frac{28}{9
   x}-\frac{29}{9}\right) \lx+\frac{26 x^2}{9}+\ly^2
   \left(\frac{x}{3}-\frac{2}{3}+\frac{2}{3 x}\right)
\brk
+\ly
   \left(\frac{x}{3}+\frac{28}{9 (1-x)}-\frac{32}{9}+\frac{46}{9
   x}\right)-\frac{26 x}{9}+\pi ^2 \left(\frac{16 x^2}{9}-\frac{16
   x}{9}-\frac{17}{18 (1-x)}+\frac{25}{9}-\frac{17}{18 x}\right)-\frac{346}{27
   (1-x)}
\brk
-\frac{346}{27 x}+\frac{578}{27}
\right\}
+
\lm^2\left[
-\frac{32 x^2}{9}+\frac{32 x}{9}+\frac{22}{9 (1-x)}-6+\frac{22}{9 x}
\right]
+
\lm \ls\left[
-\frac{64 x^2}{9}+\frac{64 x}{9}+\frac{44}{9 (1-x)}-12
\brk
+\frac{44}{9 x}
\right]
+
\ls^2\left[
-\frac{22}{9 x}+\frac{32}{9}-\frac{22}{9 (1-x)}
\right]
+
\lm\left[
-\frac{50 x^2}{27}+\frac{50 x}{27}+\pi ^2 \left(\frac{2 x^2}{3}-\frac{2
   x}{3}-\frac{1}{3 (1-x)}+1-\frac{1}{3 x}\right)
\brk
+\frac{73}{27
   (1-x)}-\frac{32}{9}+\frac{73}{27 x}
\right]
+
\ls\left[
-\frac{64 x^2}{9}+\frac{64 x}{9}+\lx \left(-\frac{44}{9
   x}+\frac{64}{9}-\frac{44}{9 (1-x)}\right)
\brk
+\ly \left(-\frac{44}{9
   x}+\frac{64}{9}-\frac{44}{9 (1-x)}\right)+\pi ^2 \left(-\frac{1}{9
   x}+\frac{2}{9}-\frac{1}{9 (1-x)}\right)+\frac{56}{27
   (1-x)}-\frac{340}{27}+\frac{56}{27 x}
\right]
\ebrk
+
\left(\left(\frac{4}{x}-2\right) \ly^2+\left(-\frac{x}{3}+\frac{2}{3
   (1-x)}+\frac{23}{3}\right) \ly+2 x^2-\frac{4 x}{9}+\frac{47}{9
   (1-x)}-\frac{3}{(x-1)^2}-\frac{49}{9}\right)
   \lx^2
\ebrk
+\left(\left(\frac{4 x}{3}-\frac{8}{3}+\frac{4}{3 x}\right)
   \ly^2+\left(-4 x^2+4 x+\pi ^2 \left(\frac{4}{3}-\frac{8}{3
   x}\right)-\frac{44}{9 (1-x)}+\frac{70}{9}-\frac{44}{9 x}\right)
   \ly+\frac{7 x}{18}
\brk
+\pi ^2 \left(\frac{7 x}{3}-\frac{7}{3
   (1-x)}-\frac{5}{9}+\frac{1}{9 x}\right)-\frac{220}{27 (1-x)}+\frac{41}{27
   x}-\frac{203}{54}\right) \lx+\frac{130 x^2}{27}+\pi ^4
   \left(\frac{11}{45}-\frac{22}{45 (1-x)}\right)
\ebrk
+\s22
   \left(\frac{16}{1-x}-\frac{16}{x}\right)+\li3 \left(\ly
   \left(8-\frac{16}{x}\right)-2 x+\frac{4}{3 (1-x)}-18\right)+\s12
   \left(\lx \left(\frac{16}{x}-8\right)
\brk
-2 x-\frac{4}{3
   x}+20\right)-\frac{130 x}{27}+\pi ^2 \left(-\frac{56 x^2}{27}+\frac{56
   x}{27}+\frac{197}{54 (1-x)}-\frac{9}{2}+\frac{197}{54 x}\right)+\ly^2
   \left(2 x^2-\frac{32 x}{9}-\frac{35}{9}
\brk
+\frac{47}{9
   x}-\frac{3}{x^2}\right)+\li2 \left(\pi ^2 \left(\frac{8}{3
   (1-x)}-\frac{8}{3 x}\right)+\lx \left(\ly
   \left(\frac{16}{x}-8\right)+2 x-\frac{4}{3 (1-x)}+18\right)\right)
\ebrk
+\left(2
   x-\frac{10}{3 (1-x)}+22-\frac{2}{x}\right) \z3+\ly \left(\pi ^2
   \left(-\frac{7 x}{3}+\frac{1}{9 (1-x)}+\frac{16}{9}-\frac{7}{3
   x}\right)-\frac{7 x}{18}+\left(\frac{16}{x}-8\right) \z3
\brk
+\frac{41}{27
   (1-x)}-\frac{220}{27 x}-\frac{91}{27}\right)+\frac{709}{81
   (1-x)}+\frac{709}{81 x}+\frac{113}{162},
\end{eqnarray}

\begin{eqnarray}
F_h &=& \frac{1}{\ep^2}\left\{
\lm\left[
\frac{8}{3 x}-\frac{16}{3}+\frac{8}{3 (1-x)}
\right]
+
\ls\left[
\frac{8}{3 x}-\frac{16}{3}+\frac{8}{3 (1-x)}
\right]
\right\}
+
\frac{1}{\ep}\left\{
\lm^2\left[
-\frac{16 x^2}{3}+\frac{16 x}{3}
\ibrk
+\frac{4}{3 (1-x)}-\frac{16}{3}+\frac{4}{3 x}
\right]
+
\lm \ls\left[
-\frac{16 x^2}{3}+\frac{16 x}{3}-\frac{8}{3 (1-x)}+\frac{8}{3}-\frac{8}{3 x}
\right]
+
\ls^2\left[
-\frac{4}{x}+8-\frac{4}{1-x}
\right]
\brk
+
\lm\left[
-\frac{16 x^2}{3}+\frac{16 x}{3}+\lx \left(-\frac{8}{3
   x}+\frac{16}{3}-\frac{8}{3 (1-x)}\right)+\ly \left(-\frac{8}{3
   x}+\frac{16}{3}-\frac{8}{3 (1-x)}\right)+\frac{20}{9
   (1-x)}
\ibrk
-\frac{112}{9}+\frac{20}{9 x}
\right]
+
\ls\left[
-\frac{16 x^2}{3}+\frac{16 x}{3}+\lx \left(-\frac{8}{3
   x}+\frac{16}{3}-\frac{8}{3 (1-x)}\right)+\ly \left(-\frac{8}{3
   x}+\frac{16}{3}-\frac{8}{3 (1-x)}\right)
\ibrk
+\frac{20}{9
   (1-x)}-\frac{112}{9}+\frac{20}{9 x}
\right]
+
\pi ^2 \left(-\frac{2}{9 x}+\frac{4}{9}-\frac{2}{9 (1-x)}\right)
\right\}
+
\lm^3\left[
\frac{52 x^2}{9}-\frac{52 x}{9}-\frac{8}{3 (1-x)}+\frac{74}{9}-\frac{8}{3 x}
\right]
\ebrk
+
\lm^2 \ls\left[
\frac{40 x^2}{3}-\frac{40 x}{3}-\frac{4}{1-x}+\frac{44}{3}-\frac{4}{x}
\right]
+
\lm \ls^2\left[
8 x^2-8 x+\frac{4}{3 (1-x)}+\frac{4}{3}+\frac{4}{3 x}
\right]
\ebrk
+
\ls^3\left[
\frac{28}{9 x}-\frac{56}{9}+\frac{28}{9 (1-x)}
\right]
+
\lm^2\left[
\frac{86 x^2}{9}-\frac{86 x}{9}+\lx
   \left(\frac{2}{x}-4+\frac{2}{1-x}\right)+\ly
   \left(\frac{2}{x}-4+\frac{2}{1-x}\right)
\brk
-\frac{25}{3
   (1-x)}+\frac{193}{9}-\frac{25}{3 x}
\right]
+
\lm \ls\left[
\frac{40 x^2}{3}-\frac{40 x}{3}+\lx \left(\frac{16}{3
   x}-\frac{32}{3}+\frac{16}{3 (1-x)}\right)
\brk
+\ly \left(\frac{16}{3
   x}-\frac{32}{3}+\frac{16}{3 (1-x)}\right)-\frac{32}{9
   (1-x)}+\frac{196}{9}-\frac{32}{9 x}
\right]
+
\ls^2\left[
8 x^2-8 x+\lx \left(\frac{4}{x}-8+\frac{4}{1-x}\right)
\brk
+\ly
   \left(\frac{4}{x}-8+\frac{4}{1-x}\right)+\frac{14}{9
   (1-x)}+\frac{80}{9}+\frac{14}{9 x}
\right]
+
\lm\left[
\left(-\frac{x}{3}+\frac{2}{3 (1-x)}-\frac{1}{3}\right)
   \lx^2+\left(\ly \left(\frac{4}{3 x}
\iibrk
-\frac{8}{3}+\frac{4}{3
   (1-x)}\right)-\frac{x}{3}+\frac{68}{9 (1-x)}+\frac{50}{9
   x}-\frac{61}{9}\right) \lx+\frac{458 x^2}{27}+\ly^2
   \left(\frac{x}{3}-\frac{2}{3}+\frac{2}{3 x}\right)
\brk
+\ly
   \left(\frac{x}{3}+\frac{50}{9 (1-x)}-\frac{64}{9}+\frac{68}{9
   x}\right)-\frac{458 x}{27}+\pi ^2 \left(\frac{28 x^2}{9}-\frac{28
   x}{9}-\frac{4}{3 (1-x)}+\frac{38}{9}-\frac{4}{3 x}\right)-\frac{713}{27
   (1-x)}
\brk
-\frac{713}{27 x}+\frac{1481}{27}
\right]
+
\ls\left[
\left(-\frac{2 x}{3}+\frac{4}{3 (1-x)}-\frac{2}{3}\right)
   \lx^2+\left(\ly \left(\frac{8}{3 x}-\frac{16}{3}+\frac{8}{3
   (1-x)}\right)-\frac{2 x}{3}+\frac{16}{3 (1-x)}
\ibrk
+\frac{4}{3
   x}+\frac{2}{3}\right) \lx-\frac{4 x^2}{3}+\ly^2 \left(\frac{2
   x}{3}-\frac{4}{3}+\frac{4}{3 x}\right)+\ly \left(\frac{2
   x}{3}+\frac{4}{3 (1-x)}+\frac{16}{3 x}\right)+\frac{4 x}{3}+\pi ^2
   \left(\frac{32 x^2}{9}-\frac{32 x}{9}
\ibrk
-\frac{14}{9
   (1-x)}+\frac{44}{9}-\frac{14}{9 x}\right)-\frac{124}{9 (1-x)}-\frac{124}{9
   x}+\frac{184}{9}
\right]
+
\left(-\frac{2 x}{9}+\frac{2}{9 (1-x)}+\frac{2}{9}-\frac{2}{9 x}\right)
   \lx^3
\ebrk
+\left(\left(\frac{4}{x}-2\right) \ly^2+8 \ly+2
   x^2-\frac{23 x}{18}+\frac{53}{9
   (1-x)}-\frac{1}{x}-\frac{3}{(x-1)^2}-\frac{35}{18}\right)
   \lx^2
\ebrk
+\left(\left(\frac{4 x}{3}+\frac{2}{3
   (1-x)}-4+\frac{2}{x}\right) \ly^2+\left(-4 x^2+4 x+\pi ^2
   \left(\frac{4}{3}-\frac{8}{3 x}\right)-\frac{20}{9
   (1-x)}+\frac{10}{9}-\frac{20}{9 x}\right) \ly
\brk
+\frac{31 x}{18}+\pi ^2
   \left(2 x-\frac{20}{9 (1-x)}+\frac{2}{9}-\frac{4}{9 x}\right)-\frac{50}{27
   (1-x)}+\frac{229}{27 x}-\frac{709}{54}\right) \lx+\frac{1091
   x^2}{27}
\ebrk
+\pi ^4 \left(\frac{11}{45}-\frac{22}{45 (1-x)}\right)+\s22
   \left(\frac{16}{1-x}-\frac{16}{x}\right)+\s12 \left(\lx
   \left(\frac{16}{x}-8\right)-\frac{8 x}{3}-\frac{4}{3
   (1-x)}-\frac{4}{x}+24\right)
\ebrk
+\li3 \left(\ly
   \left(8-\frac{16}{x}\right)-\frac{8 x}{3}+\frac{4}{1-x}+\frac{4}{3
   x}-\frac{64}{3}\right)+\ly^3 \left(\frac{2 x}{9}-\frac{2}{9
   (1-x)}+\frac{2}{9 x}\right)-\frac{1091 x}{27}
\ebrk
+\pi ^2 \left(-\frac{22
   x^2}{9}+\frac{22 x}{9}+\frac{53}{27 (1-x)}+\frac{26}{27}+\frac{53}{27
   x}\right)+\ly^2 \left(2 x^2-\frac{49
   x}{18}-\frac{1}{1-x}-\frac{11}{9}+\frac{53}{9
   x}-\frac{3}{x^2}\right)
\ebrk
+\li2 \left(\pi ^2 \left(\frac{8}{3
   (1-x)}-\frac{8}{3 x}\right)+\lx \left(\ly
   \left(\frac{16}{x}-8\right)+\frac{8 x}{3}-\frac{4}{1-x}-\frac{4}{3
   x}+\frac{64}{3}\right)\right)+\left(\frac{8 x}{3}+\frac{20}{9
   (1-x)}
\brk
+\frac{80}{9}+\frac{44}{9 x}\right) \z3+\ly \left(\pi ^2
   \left(-2 x-\frac{4}{9 (1-x)}+\frac{20}{9}-\frac{20}{9 x}\right)-\frac{31
   x}{18}+\left(\frac{16}{x}-8\right) \z3+\frac{229}{27
   (1-x)}-\frac{50}{27 x}
\brk
-\frac{308}{27}\right)-\frac{4243}{162
   (1-x)}-\frac{4243}{162 x}+\frac{5821}{81},
\end{eqnarray}

\begin{eqnarray}
G_l &=& \frac{1}{\ep^2}\left\{
\lm\left[
-\frac{1}{x}+2-\frac{1}{1-x}
\right]
-\frac{1}{x}+2-\frac{1}{1-x}
\right\}
+
\frac{1}{\ep}\left\{
\lm^2\left[
\frac{1}{3 x}-\frac{2}{3}+\frac{1}{3 (1-x)}
\right]
\brk
+
\lm \ls\left[
\frac{2}{3 x}-\frac{4}{3}+\frac{2}{3 (1-x)}
\right]
+
\lm\left[
\frac{20}{9 x}-\frac{22}{9}+\frac{20}{9 (1-x)}
\right]
+
\ls\left[
\frac{2}{3 x}-\frac{4}{3}+\frac{2}{3 (1-x)}
\right]
\brk
+
\left(\frac{1}{3 x}-\frac{1}{3}+\frac{2}{3 (1-x)}\right)
   \lx^2+\left(\frac{1}{x}-\frac{1}{3}\right) \lx+\ly
   \left(\frac{1}{1-x}-\frac{1}{3}\right)+\pi ^2 \left(\frac{4}{9
   x}-\frac{8}{9}+\frac{4}{9 (1-x)}\right)
\brk
+\ly^2 \left(\frac{2}{3
   x}-\frac{1}{3}+\frac{1}{3 (1-x)}\right)+\frac{37}{18 (1-x)}+\frac{37}{18
   x}-\frac{19}{9}
\right\}
+
\lm^2\left[
-\frac{11}{9 x}+\frac{16}{9}-\frac{11}{9 (1-x)}
\right]
\ebrk
+
\lm \ls\left[
-\frac{22}{9 x}+\frac{32}{9}-\frac{22}{9 (1-x)}
\right]
+
\lm\left[
\pi ^2 \left(\frac{1}{6 x}-\frac{1}{3}+\frac{1}{6 (1-x)}\right)-\frac{73}{54
   (1-x)}-\frac{73}{54 x}+\frac{7}{27}
\right]
\ebrk
+
\ls\left[
-\frac{22}{9 x}+\frac{32}{9}-\frac{22}{9 (1-x)}
\right]
+
\left(\left(6-\frac{12}{x}\right) \ly^2+\left(\frac{1}{3
   x}-\frac{73}{3}+\frac{2}{3 (1-x)}\right) \ly-\frac{241}{9
   (1-x)}-\frac{11}{9 x}
\brk
+\frac{9}{(x-1)^2}+\frac{203}{9}\right)
   \lx^2+\left(\left(-4 x+8-\frac{4}{x}\right) \ly^2+\left(\pi ^2
   \left(\frac{8}{x}-4\right)-2\right) \ly+\pi ^2 \left(-\frac{20
   x}{3}+\frac{58}{9 (1-x)}
\ibrk
+\frac{13}{9}-\frac{1}{9 x}\right)-16
   x+\frac{18}{1-x}-\frac{13}{6 x}-\frac{17}{18}\right) \lx+\li2
   \left(\pi ^2 \left(\frac{8}{x}-\frac{8}{1-x}\right)+\lx
   \left(\ly \left(24-\frac{48}{x}\right)-8 x
\ibrk
+\frac{28}{3
   (1-x)}+\frac{2}{3 x}-\frac{170}{3}\right)\right)+\pi ^4 \left(\frac{22}{15
   (1-x)}-\frac{11}{15}\right)+\ly^2 \left(-\frac{241}{9
   x}+\frac{9}{x^2}+\frac{203}{9}-\frac{11}{9 (1-x)}\right)
\ebrk
+\pi ^2
   \left(-\frac{44}{27 x}-\frac{275}{27}-\frac{44}{27 (1-x)}\right)+\s22
   \left(\frac{48}{x}-\frac{48}{1-x}\right)+\li3 \left(\ly
   \left(\frac{48}{x}-24\right)+8 x
\brk
-\frac{28}{3 (1-x)}-\frac{2}{3
   x}+\frac{170}{3}\right)+\s12 \left(\lx
   \left(24-\frac{48}{x}\right)+8 x+\frac{2}{3 (1-x)}+\frac{28}{3
   x}-\frac{194}{3}\right)+\left(-8 x
\brk
+\frac{28}{3
   (1-x)}-\frac{170}{3}+\frac{2}{3 x}\right) \z3+\ly \left(16
   x+\pi ^2 \left(\frac{20 x}{3}-\frac{1}{9 (1-x)}-\frac{47}{9}+\frac{58}{9
   x}\right)+\left(24-\frac{48}{x}\right) \z3
\brk
-\frac{13}{6
   (1-x)}-\frac{305}{18}+\frac{18}{x}\right)+\frac{23}{54 (1-x)}+\frac{23}{54
   x}-\frac{80}{27},
\end{eqnarray}

\begin{eqnarray}
G_h &=& \frac{1}{\ep}\left\{
\lm^2\left[
-\frac{4}{3 x}+\frac{8}{3}-\frac{4}{3 (1-x)}
\right]
+
\lm \ls\left[
-\frac{4}{3 x}+\frac{8}{3}-\frac{4}{3 (1-x)}
\right]
+
\lm\left[
-\frac{4}{3 x}+\frac{8}{3}-\frac{4}{3 (1-x)}
\right]
\brk
+
\ls\left[
-\frac{4}{3 x}+\frac{8}{3}-\frac{4}{3 (1-x)}
\right]
\right\}
+
\lm^3\left[
\frac{13}{9 x}-\frac{26}{9}+\frac{13}{9 (1-x)}
\right]
+
\lm^2 \ls\left[
\frac{10}{3 x}-\frac{20}{3}+\frac{10}{3 (1-x)}
\right]
\ebrk
+
\lm \ls^2\left[
\frac{2}{x}-4+\frac{2}{1-x}
\right]
+
\lm^2\left[
\frac{67}{18 x}-\frac{43}{9}+\frac{67}{18 (1-x)}
\right]
+
\lm \ls\left[
\frac{14}{3 x}-\frac{20}{3}+\frac{14}{3 (1-x)}
\right]
\ebrk
+
\ls^2\left[
\frac{2}{x}-4+\frac{2}{1-x}
\right]
+
\lm\left[
\left(\frac{1}{3 x}-\frac{1}{3}+\frac{2}{3 (1-x)}\right)
   \lx^2+\left(\frac{1}{x}-\frac{1}{3}\right) \lx+\ly
   \left(\frac{1}{1-x}-\frac{1}{3}\right)
\brk
+\ly^2 \left(\frac{2}{3
   x}-\frac{1}{3}+\frac{1}{3 (1-x)}\right)+\pi ^2 \left(\frac{7}{9
   x}-\frac{14}{9}+\frac{7}{9 (1-x)}\right)+\frac{337}{54 (1-x)}+\frac{337}{54
   x}-\frac{265}{27}
\right]
\ebrk
+
\ls\left[
\left(\frac{2}{3 x}-\frac{2}{3}+\frac{4}{3 (1-x)}\right)
   \lx^2+\left(\frac{2}{x}-\frac{2}{3}\right) \lx+\ly
   \left(\frac{2}{1-x}-\frac{2}{3}\right)+\pi ^2 \left(\frac{8}{9
   x}-\frac{16}{9}+\frac{8}{9 (1-x)}\right)
\brk
+\ly^2 \left(\frac{4}{3
   x}-\frac{2}{3}+\frac{2}{3 (1-x)}\right)+\frac{5}{3 (1-x)}+\frac{5}{3
   x}-\frac{2}{3}
\right]
+
\left(\frac{2}{9 x}-\frac{2}{9}+\frac{4}{9 (1-x)}\right)
   \lx^3+\left(\left(6-\frac{12}{x}\right) \ly^2
\brk
-24
   \ly-\frac{235}{9 (1-x)}-\frac{1}{18
   x}+\frac{9}{(x-1)^2}+\frac{391}{18}\right) \lx^2+\left(\left(-4
   x+8-\frac{4}{x}\right) \ly^2+\left(\pi ^2
   \left(\frac{8}{x}-4\right)-2\right) \ly
\brk
+\pi ^2 \left(-\frac{20
   x}{3}+\frac{64}{9 (1-x)}+\frac{10}{9}+\frac{2}{9 x}\right)-16
   x+\frac{18}{1-x}-\frac{5}{2 x}+\frac{7}{18}\right) \lx+\li2
   \left(\pi ^2 \left(\frac{8}{x}-\frac{8}{1-x}\right)
\brk
+\lx
   \left(\ly \left(24-\frac{48}{x}\right)-8
   x+\frac{8}{1-x}-56\right)\right)+\pi ^4 \left(\frac{22}{15
   (1-x)}-\frac{11}{15}\right)+\ly^2 \left(-\frac{235}{9
   x}+\frac{9}{x^2}+\frac{391}{18}
\brk
-\frac{1}{18 (1-x)}\right)+\pi ^2
   \left(-\frac{7}{9 x}-\frac{31}{3}-\frac{7}{9 (1-x)}\right)+\ly^3
   \left(\frac{4}{9 x}-\frac{2}{9}+\frac{2}{9 (1-x)}\right)+\s22
   \left(\frac{48}{x}-\frac{48}{1-x}\right)
\ebrk
+\li3 \left(\ly
   \left(\frac{48}{x}-24\right)+8 x-\frac{8}{1-x}+56\right)+\s12
   \left(\lx \left(24-\frac{48}{x}\right)+8
   x+\frac{8}{x}-64\right)
\ebrk
+\left(-8 x+\frac{8}{1-x}-56\right)
   \z3+\ly \left(16 x+\pi ^2 \left(\frac{20 x}{3}+\frac{2}{9
   (1-x)}-\frac{50}{9}+\frac{64}{9 x}\right)+\left(24-\frac{48}{x}\right)
   \z3
\brk
-\frac{5}{2
   (1-x)}-\frac{281}{18}+\frac{18}{x}\right)+\frac{1013}{108
   (1-x)}+\frac{1013}{108 x}-\frac{1013}{54},
\end{eqnarray}

\begin{eqnarray}
H_l &=& \frac{1}{\ep^2}\left\{
-\frac{16 x^2}{9}+\frac{16 x}{9}+\frac{4}{9 (1-x)}-\frac{16}{9}+\frac{4}{9 x}
\right\}
+
\frac{1}{\ep}\left\{
\frac{16 x^2}{9}-\frac{16 x}{9}-\frac{8}{9 (1-x)}+\frac{8}{3}-\frac{8}{9 x}
\right\}
-\frac{8 x^2}{9}+\frac{8 x}{9}
\ebrk
+\frac{4}{9 (1-x)}-\frac{10}{9}+\frac{4}{9 x},
\end{eqnarray}

\begin{eqnarray}
H_{lh} &=& \frac{1}{\ep}\left\{
\lm\left[
-\frac{32 x^2}{9}+\frac{32 x}{9}+\frac{8}{9 (1-x)}-\frac{32}{9}+\frac{8}{9 x}
\right]
+
\ls\left[
-\frac{32 x^2}{9}+\frac{32 x}{9}+\frac{8}{9 (1-x)}-\frac{32}{9}+\frac{8}{9 x}
\right]
\right\}
\ebrk
+
\lm^2\left[
\frac{16 x^2}{9}-\frac{16 x}{9}-\frac{4}{9 (1-x)}+\frac{16}{9}-\frac{4}{9 x}
\right]
+
\lm \ls\left[
\frac{32 x^2}{9}-\frac{32 x}{9}-\frac{8}{9 (1-x)}+\frac{32}{9}-\frac{8}{9 x}
\right]
\ebrk
+
\ls^2\left[
\frac{16 x^2}{9}-\frac{16 x}{9}-\frac{4}{9 (1-x)}+\frac{16}{9}-\frac{4}{9 x}
\right]
+
\lm\left[
\frac{32 x^2}{9}-\frac{32 x}{9}-\frac{16}{9 (1-x)}+\frac{16}{3}-\frac{16}{9 x}
\right]
\ebrk
+
\ls\left[
\frac{32 x^2}{9}-\frac{32 x}{9}-\frac{16}{9 (1-x)}+\frac{16}{3}-\frac{16}{9 x}
\right]
-\frac{8 x^2}{9}+\frac{8 x}{9}+\pi ^2 \left(\frac{8 x^2}{27}-\frac{8
   x}{27}-\frac{2}{27 (1-x)}+\frac{8}{27}-\frac{2}{27 x}\right)
\ebrk
-\frac{2}{9},
\end{eqnarray}

\begin{eqnarray}
H_h &=& \lm^2\left[
-\frac{16 x^2}{9}+\frac{16 x}{9}+\frac{4}{9 (1-x)}-\frac{16}{9}+\frac{4}{9 x}
\right]
+
\lm \ls\left[
-\frac{32 x^2}{9}+\frac{32 x}{9}+\frac{8}{9 (1-x)}-\frac{32}{9}+\frac{8}{9 x}
\right]
\ebrk
+
\ls^2\left[
-\frac{16 x^2}{9}+\frac{16 x}{9}+\frac{4}{9 (1-x)}-\frac{16}{9}+\frac{4}{9 x}
\right],
\end{eqnarray}

\begin{eqnarray}
I_l &=& \frac{1}{\ep^2}\left\{
-\frac{4}{9 x}+\frac{8}{9}-\frac{4}{9 (1-x)}
\right\}
+
\frac{1}{\ep}\left\{
\frac{8}{9 x}-\frac{8}{9}+\frac{8}{9 (1-x)}
\right\}
-\frac{4}{9 x}-\frac{4}{9 (1-x)},
\end{eqnarray}

\begin{eqnarray}
I_{lh} &=& \frac{1}{\ep}\left\{
\lm\left[
-\frac{8}{9 x}+\frac{16}{9}-\frac{8}{9 (1-x)}
\right]
+
\ls\left[
-\frac{8}{9 x}+\frac{16}{9}-\frac{8}{9 (1-x)}
\right]
\right\}
+
\lm^2\left[
\frac{4}{9 x}-\frac{8}{9}+\frac{4}{9 (1-x)}
\right]
\ebrk
+
\lm \ls\left[
\frac{8}{9 x}-\frac{16}{9}+\frac{8}{9 (1-x)}
\right]
+
\ls^2\left[
\frac{4}{9 x}-\frac{8}{9}+\frac{4}{9 (1-x)}
\right]
+
\lm\left[
\frac{16}{9 x}-\frac{16}{9}+\frac{16}{9 (1-x)}
\right]
\ebrk
+
\ls\left[
\frac{16}{9 x}-\frac{16}{9}+\frac{16}{9 (1-x)}
\right]
+
\pi ^2 \left(\frac{2}{27 x}-\frac{4}{27}+\frac{2}{27 (1-x)}\right),
\end{eqnarray}

\begin{eqnarray}
I_h &=& \lm^2\left[
-\frac{4}{9 x}+\frac{8}{9}-\frac{4}{9 (1-x)}
\right]
+
\lm \ls\left[
-\frac{8}{9 x}+\frac{16}{9}-\frac{8}{9 (1-x)}
\right]
+
\ls^2\left[
-\frac{4}{9 x}+\frac{8}{9}-\frac{4}{9 (1-x)}
\right].
\ebrk
\end{eqnarray}
%
}}

Notice that apart from the classic polylogarithms up to weight four,
our results are also expressed in terms of Nielsen polylogarithms
\begin{equation}
\mathrm{S}_{n,p}(x) = \frac{(-1)^{n+p-1}}{(n-1)!p!} \int_0^1 dy
\frac{\log^{n-1}(y)\log^p(1-x y)}{y}.
\end{equation}
%

%
%
\section{Conclusions}
\label{sec:conclusions}

In the present article we have computed the two-loop virtual QCD corrections to the production of heavy-quarks
for the gluon fusion process in the ultra-relativistic limit.
Previously we had already obtained the corresponding results for quark-quark scattering~\cite{Czakon:2007ej}.
Taken together, these results complete the two-loop radiative QCD corrections to heavy-quark production in hadron-hadron collisions
in the limit when all kinematical invariants are large compared to the heavy-quark mass.
The remaining channel with a $gq$ (or $g{\bar q}$) initial state involves at most one-loop corrections
in a consistent NNLO treatment. At one-loop the complete structure of the singularities as well as all large logarithms
in the heavy-quark mass $m$ can therefore be entirely treated with the methods of Ref.~\cite{Catani:2000ef}.

Our derivation relies on the combination of two completely different methods
and we have ensured substantial overlap between them.
In this way we have had mutual and highly non-trivial cross checks on our direct calculation of massive Feynman diagrams and
on the QCD factorization approach~\cite{Mitov:2006xs}.
In particular through the latter method we have had the possibility to relate to various
massless~\cite{Anastasiou:2001sv,Bern:2003ck,Glover:2003cm,Moch:2005id,Gehrmann:2005pd}
and massive results~\cite{Bernreuther:2004ih} available in the literature and we have
found consistency.

The results for gluon fusion in the present article (and Ref.~\cite{Czakon:2007ej} for quark-quark scattering) are
of direct relevance whenever power corrections in the heavy-quark mass are negligible.
This is certainly the case for hadro-production of bottom pairs over a large kinematical range at colliders
and, to a lesser extent perhaps, for $t{\bar t}$-production at LHC energies.
Here, possible improvements would come from the systematic computation of power corrections
in the heavy-quark mass, i.e. terms proportional to $(m^2/s)^k$ with $k\ge 1$
to improve the convergence of the small-mass expansion.
This could be achieved, for instance, by extending the methods of Sec.~\ref{sec:method}
for the direct calculation of massive Feynman diagrams to higher powers in $m$.

Finally, it is clear, that our result for $\langle {\cal M}^{(0)} | {\cal M}^{(2)} \rangle $
still has to be combined with the tree-level $2 \to 4$, the one-loop $2 \to 3$
as well as the square of the one-loop $2 \to 2$ processes $\langle {\cal M}^{(1)} | {\cal M}^{(1)} \rangle $
in order to yield physical cross sections.
Some of the matrix elements including the full mass dependence can be easily generated,
others have become available in the literature only rather recently,
see e.g. Refs.\cite{Korner:2005rg,Dittmaier:2007wz}.
The combination of all these contributions enables the analytic cancellation of the remaining
infrared divergences as well as the isolation of the initial state singularities.
The latter will have to be absorbed into parton distribution functions of, say, the proton
in order to match with a precise parton evolution at NNLO~\cite{Moch:2004pa,Vogt:2004mw}.
All these remaining steps are necessary prerequisites e.g. to the construction of numerical programs
which then provide NNLO QCD estimates of observable scattering cross sections for
heavy-quark hadro-production.

A {\sc Mathematica} file with our results can be obtained by downloading the
source from the preprint server {\tt http://arXiv.org}.
The results are also available from the authors upon request.

%
%
{\bf{Acknowledgments:}}
We are grateful to C.~Anastasiou, Z.~Bern, L.~Dixon and N.~Glover for communication on the results
of Refs.~\cite{Anastasiou:2001sv,Bern:2003ck,Glover:2003cm,DeFreitas:2004tk}
and for useful discussions. We would also like to thank V.A.Smirnov for an
interesting exchange of opinions on MB representations of non-planar graphs.
M.C. and A.M. thank the Alexander von Humboldt Foundation for support
through a Sofja Kovalevskaja Award and a research grant, respectively.
S.M. acknowledges contract VH-NG-105 by the Helmholtz Gemeinschaft.
This work was also partially supported by the Deutsche Forschungsgemeinschaft
in Sonderforschungs\-be\-reich/Transregio~9.

%
%
\appendix
\renewcommand{\theequation}{\ref{sec:appA}.\arabic{equation}}
\setcounter{equation}{0}
\section{Asymptotics of the massive non-planar scalar integral}
\label{sec:appA}

Here, we give the leading high energy behavior of the non-planar scalar integral
with a massive loop corresponding to the MB representation Eq.~(\ref{rep1})
{\small{
\begin{eqnarray}
&&s^{3+2\ep} \; I_{NP} \; = \;
\frac{\pi^2}{\sqrt{\frac{m^2}{s}\; x(1-x)}}\left\{-\pi^2 + i \pi \left[
    -{\lm}+{\lx}+{\ly}+4 \log (2) \right]\right\}
\ebrk
+
{\lm}^4\left[
\frac{7}{12 x}+\frac{7}{12 (1-x)}
\right]
+
{\lm}^3\left[
{\ly} \left(-\frac{7}{3 x}-\frac{5}{3 (1-x)}\right)+{\lx}
   \left(-\frac{5}{3 x}-\frac{7}{3 (1-x)}\right)-\frac{4}{3 (1-x)}-\frac{4}{3
   x}
\right]
\ebrk
+
{\lm}^2\left[
\left(\frac{3}{2 x}+\frac{3}{2 (1-x)}\right) {\lx}^2+\left({\ly}
   \left(\frac{4}{x}+\frac{4}{1-x}\right)-\frac{6}{1-x}+\frac{4}{x}\right)
   {\lx}+{\ly} \left(\frac{4}{1-x}-\frac{6}{x}\right)
\brk
+{\ly}^2
   \left(\frac{3}{2 x}+\frac{3}{2 (1-x)}\right)+\pi ^2
   \left(\frac{2}{x}+\frac{2}{1-x}\right)+\frac{8}{1-x}+\frac{8}{x}
\right]
+
{\lm}\left[
\left(\frac{1}{3 (1-x)}-\frac{1}{3 x}\right) {\lx}^3
\brk
+\left({\ly}
   \left(-\frac{1}{x}-\frac{3}{1-x}\right)-\frac{4}{x}\right)
   {\lx}^2+\left(\left(-\frac{3}{x}-\frac{1}{1-x}\right)
   {\ly}^2+\left(\frac{12}{x}+\frac{12}{1-x}\right) {\ly}+\pi ^2
   \left(\frac{2}{1-x}-\frac{10}{3
   x}\right)
\ibrk
+\frac{16}{1-x}-\frac{16}{x}\right) {\lx}+\pi ^2
   \left(-\frac{4}{3 x}-\frac{4}{3 (1-x)}\right)+{\ly}^3 \left(\frac{1}{3
   x}-\frac{1}{3 (1-x)}\right)+{\ly} \left(\pi ^2
   \left(\frac{2}{x}-\frac{10}{3
   (1-x)}\right)
\ibrk
-\frac{16}{1-x}+\frac{16}{x}\right)+\left(\frac{20}{x}+\frac{20}{1-x}\right)
   {\z3}-\frac{4
   {\ly}^2}{1-x}-\frac{32}{1-x}-\frac{32}{x}
\right]
+
\left(-\frac{1}{12 x}-\frac{5}{12 (1-x)}\right) {\lx}^4
\ebrk
+\left({\ly}
   \left(\frac{2}{3 (1-x)}-\frac{2}{3 x}\right)+\frac{2}{1-x}+\frac{4}{3
   x}\right) {\lx}^3+\left(\left(\frac{1}{2 x}+\frac{3}{2 (1-x)}\right)
   {\ly}^2+\frac{2 {\ly}}{x}+\pi ^2 \left(\frac{4}{3
   x}-\frac{1}{1-x}\right)
\brk
+\frac{8}{x}\right)
   {\lx}^2+\left(\left(\frac{2}{3 x}-\frac{2}{3 (1-x)}\right)
   {\ly}^3+\left(-\frac{6}{x}-\frac{6}{1-x}\right) {\ly}^2+\left(\pi
   ^2 \left(\frac{1}{1-x}-\frac{11}{3
   x}\right)-\frac{16}{1-x}-\frac{16}{x}\right) {\ly}
\brk
+\pi ^2
   \left(\frac{4}{3
   x}+\frac{6}{1-x}\right)+\left(\frac{10}{1-x}-\frac{18}{x}\right)
   {\z3}-\frac{16}{1-x}+\frac{32}{x}\right) {\lx}+{\li2}
   \left(\left(\frac{2}{1-x}-\frac{4}{x}\right) {\lx}^2
\brk
+\left({\ly}
   \left(\frac{6}{1-x}-\frac{6}{x}\right)+\frac{12}{1-x}+\frac{16}{x}\right)
   {\lx}+\pi ^2 \left(\frac{14}{3 (1-x)}-\frac{14}{3
   x}\right)\right)+{\ly}^2 \left(\pi ^2 \left(\frac{4}{3
   (1-x)}-\frac{1}{x}\right)
\brk
+\frac{8}{1-x}\right)+{\li4}
   \left(\frac{12}{1-x}-\frac{24}{x}\right)+{\li3} \left({\ly}
   \left(\frac{6}{x}-\frac{6}{1-x}\right)+{\lx}
   \left(\frac{16}{x}-\frac{8}{1-x}\right)-\frac{12}{1-x}-\frac{16}{x}\right)
\ebrk
+
   {\s13} \left(\frac{24}{1-x}-\frac{12}{x}\right)+{\ly}^4
   \left(-\frac{5}{12 x}-\frac{1}{12 (1-x)}\right)+\pi ^4 \left(\frac{47}{180
   x}-\frac{31}{36 (1-x)}\right)+{\ly}^3 \left(\frac{2}{x}
\brk
+\frac{4}{3
   (1-x)}\right)+\pi ^2 \left(\frac{4}{x}+\frac{4}{1-x}\right)+{\s12}
   \left({\lx} \left(\frac{14}{1-x}-\frac{14}{x}\right)+{\ly}
   \left(\frac{8}{1-x}-\frac{4}{x}\right)+\frac{16}{1-x}+\frac{12}{x}\right)
\ebrk
+{\s22}
   \left(\frac{20}{x}-\frac{20}{1-x}\right)+\left(-\frac{4}{x}-\frac{8}{1-x}\right)
   {\z3}+{\ly} \left(\pi ^2 \left(\frac{6}{x}+\frac{4}{3
   (1-x)}\right)+\left(\frac{8}{x}-\frac{20}{1-x}\right)
   {\z3}+\frac{32}{1-x}
\brk
-\frac{16}{x}\right)+\frac{64}{1-x}+\frac{64}{x}
\ebrk
+
i \pi \left\{
{\lm}^3\left[
-\frac{5}{3 x}-\frac{5}{3 (1-x)}
\right]
+
{\lm}^2\left[
\frac{2 {\lx}}{x}+\frac{2 {\ly}}{1-x}-\frac{6}{1-x}-\frac{6}{x}
\right]
+
{\lm}\left[
\left(\frac{1}{x}+\frac{1}{1-x}\right) {\lx}^2+\frac{12
   {\lx}}{x}
\ibrk
+{\ly}^2 \left(\frac{1}{x}+\frac{1}{1-x}\right)+\pi ^2
   \left(\frac{10}{3 x}+\frac{10}{3 (1-x)}\right)+\frac{12
   {\ly}}{1-x}+\frac{16}{1-x}+\frac{16}{x}
\right]
+
\left(-\frac{4}{3 x}-\frac{2}{3 (1-x)}\right) {\lx}^3
\brk
+\left({\ly}
   \left(\frac{1}{x}-\frac{1}{1-x}\right)-\frac{6}{x}\right)
   {\lx}^2+\left(\frac{4 {\ly}}{x}-\frac{10 \pi ^2}{3
   x}-\frac{16}{x}\right) {\lx}+{\ly} \left(\pi ^2 \left(\frac{1}{3
   x}-\frac{3}{1-x}\right)-\frac{16}{1-x}\right)
\brk
+{\s12}
   \left(-\frac{6}{x}-\frac{2}{1-x}\right)+{\ly}^3 \left(-\frac{2}{3
   x}-\frac{4}{3 (1-x)}\right)+{\li3}
   \left(\frac{2}{x}+\frac{6}{1-x}\right)+{\li2} \left(-\frac{4
   {\lx}}{1-x}
\ibrk
+{\ly}
   \left(-\frac{2}{x}-\frac{2}{1-x}\right)-\frac{4}{1-x}+\frac{4}{x}\right)+
   \left(\frac{18}{x}+\frac{14}{1-x}\right) {\z3}-\frac{6
   {\ly}^2}{1-x}+\frac{2 \pi ^2}{3 (1-x)}-\frac{16}{1-x}-\frac{16}{x}
\right\}
\ebrk
+{\cal O}\left(\sqrt{\frac{m^2}{s}}\right)
\, .
\end{eqnarray}
}}

%
%

\end{document}